\documentclass[notitlepage,a4paper,aps,prd,onecolumn,superscriptaddress,nofootinbib,groupedaddress]{revtex4}
\usepackage{amsmath}
\usepackage{amsfonts,color}
\usepackage{amsmath}
\usepackage{amsthm}
\newtheorem{theorem}{Theorem}
\usepackage{pdfpages}
\newcommand{\dd}{{\rm d}}
\usepackage{amsfonts,color}
\usepackage{amssymb,float}

\newcommand{\lc}[1]{\accentset{\circ}{#1}}

\usepackage{amsmath,accents}

\usepackage[utf8]{inputenc}
\allowdisplaybreaks
\usepackage{color}
\usepackage{hyperref}
\hypersetup{
    colorlinks=true,
    linkcolor=blue,
    filecolor=magenta,      
   citecolor=blue
}

\usepackage{enumitem}

\usepackage{tikz}
\usetikzlibrary{shapes.geometric}
\usetikzlibrary{arrows.meta,arrows}

\begin{document}
\title{Black hole solutions in scalar-tensor symmetric teleparallel gravity}

\author{Sebastian Bahamonde}
\email{bahamonde.s.aa@m.titech.ac.jp}
\affiliation{Department of Physics, Tokyo Institute of Technology
1-12-1 Ookayama, Meguro-ku, Tokyo 152-8551, Japan.}
\affiliation{Laboratory of Theoretical Physics, Institute of Physics, University of Tartu, W. Ostwaldi 1, 50411 Tartu, Estonia.}

\author{Jorge Gigante Valcarcel}
\email{jorge.gigante.valcarcel@ut.ee}
\affiliation{Laboratory of Theoretical Physics, Institute of Physics, University of Tartu, W. Ostwaldi 1, 50411 Tartu, Estonia.}

\author{Laur J\"arv}
\email{laur.jarv@ut.ee}
\affiliation{Laboratory of Theoretical Physics, Institute of Physics, University of Tartu, W. Ostwaldi 1, 50411 Tartu, Estonia.}

\author{Joosep Lember}

\affiliation{Laboratory of Theoretical Physics, Institute of Physics, University of Tartu, W. Ostwaldi 1, 50411 Tartu, Estonia.}
\begin{abstract}
Symmetric teleparallel gravity is constructed with a nonzero nonmetricity tensor while both torsion and curvature are vanishing. In this framework, we find exact scalarised spherically symmetric static solutions in scalar-tensor theories built with a nonminimal coupling between the nonmetricity scalar and a scalar field. It turns out that the Bocharova-Bronnikov-Melnikov-Bekenstein solution has a symmetric teleparallel analogue (in addition to the recently found metric teleparallel analogue), while some other of these solutions describe scalarised black hole configurations that are not known in the Riemannian or metric teleparallel scalar-tensor case. To aid the analysis we also derive no-hair theorems for the theory. Since the symmetric teleparallel scalar-tensor models also include $f(Q)$ gravity, we shortly discuss this case and further prove a theorem which says that by imposing that the metric functions are the reciprocal of each other ($g_{rr}=1/g_{tt}$), the $f(Q)$ gravity theory reduces to the symmetric teleparallel equivalent of general relativity (plus a cosmological constant), and the metric takes the (Anti)de-Sitter–Schwarzschild form.
\end{abstract}

\maketitle

\section{Introduction}

One of the celebrated results of classic general relativity (GR) is the so called \textit{no-hair} theorem \cite{Israel:1967wq,Israel:1967za,Carter:1971zc}, which says that the solutions of the Einstein-Maxwell vacuum system are uniquely determined by the mass, charge, and angular momentum of the configuration. However, after including to the theory some other fields the situation becomes more intriguing. In some cases the application range of the theorems can be enlarged \cite{Bekenstein:1972ny,Bekenstein:1971hc,Bekenstein:1972ky,Heusler:1992ss,Bekenstein:1995un,Sudarsky:1995zg,Zannias:1994jf} but ``hairy'' solutions also appear, namely nontrivial configurations with e.g.\ extra scalar fields \cite{Droz:1991cx,Achucarro:1995nu,Anabalon:2013qua,Cadoni:2015gfa,Herdeiro:2014goa,Herdeiro:2015gia} or non-Abelian vector fields \cite{Bartnik:1988am,Volkov:1998cc,Bizon:1990sr}. The extra fields can modify the geometry despite the absence of corresponding charges derived from a Gauss law. In the scalar-tensor theories, where the scalar field is nonminimally coupled to the curvature scalar in the action and starts to participate in mediating the gravitational interaction, the no-hair theorems can still be formulated under certain assumptions \cite{Hawking:1972qk,Saa:1996qq,Mayo:1996mv,Sotiriou:2011dz,Graham:2014ina}, yet interesting examples of analytic solutions dressed with scalar ``hair'' are also known \cite{bocharova1970exact,Bekenstein:1974sf,Bekenstein:1975ts,Bronnikov:1978mx,Nucamendi:1995ex,Astorino:2014mda}. See the Refs.\ \cite{Chrusciel:2012jk,Herdeiro:2015waa,Faraoni:2021nhi} for  informative reviews on the topic. Scalar-tensor and other extensions of GR can be motivated by various theoretical and phenomenological reasons~\cite{Clifton:2011jh,Capozziello:2011et,Nojiri:2010wj,Nojiri:2017ncd,Joyce:2014kja,CANTATA:2021ktz}, thus checking the repertoire of black hole solutions in these theories with the fresh observational data, like the recently reported black hole images \cite{Psaltis:2018xkc,EventHorizonTelescope:2022xqj,Vagnozzi:2022moj} as well as the gravitational waves from binary black hole coalescences \cite{LIGOScientific:2019fpa,LIGOScientific:2021sio} is very much the call of the day.

From a deeper theoretical point of view, in general relativity the gravitational interaction is characterised by a metric tensor and a symmetric metric-compatible affine connection, which can be properly introduced under the projective invariance of the Einstein-Hilbert action~\cite{Dadhich:2010xa,Afonso:2017bxr}. Consequently, according to the Fundamental Theorem
of Riemannian geometry, the metric tensor determines the form of the affine connection as the Levi-Civita connection, and the gravitational field is fully ascribed to the curvature tensor. Nevertheless, the consideration of a general affine connection with torsion and nonmetricity has shown to present equivalent formulations of the gravitational phenomena in the framework of post-Riemannian geometry~\cite{BeltranJimenez:2018vdo,BeltranJimenez:2019tjy}.
In particular, the notion of teleparallelism allows GR to be recasted within an affinely connected metric space-time equipped with either a connection characterised by nonzero torsion but vanishing curvature and nonmetricity, yielding the Teleparallel Equivalent of General Relativity (TEGR)~\cite{Aldrovandi:2013wha,Krssak:2018ywd}, or with
a nonmetric-compatible connection with vanishing curvature and torsion, leading to the framework of Symmetric Teleparallel Equivalent of General Relativity (STEGR)~\cite{Nester:1998mp}. Then, a crucial geometric identity for the construction of such theories relates the Levi-Civita Ricci scalar $\lc{R}$ present in the Einstein-Hilbert action of GR to the torsion scalar $T$ and a total divergence of the vectorial part of the torsion tensor, or the nonmetricity scalar $Q$ and a total divergence of the difference between the two independent traces of the nonmetricity tensor, respectively. The total divergence constitutes a respective boundary term $B$, which does not affect the resulting field equations.
This guarantees that the solutions of GR are reproduced in TEGR as well as STEGR, albeit in different geometric setups with extra freedom located in the post-Riemannian part of the connection.

Like the original purely metric based formulation of general relativity, the teleparallel theories can be extended by considering a gravitational action depending on a general function of the ruling geometric quantity like $f(T)$ \cite{Bengochea:2008gz,Linder:2010py} or $f(Q)$~\cite{BeltranJimenez:2017tkd} gravity, or by introducing a nonminimal coupling between a scalar field and this quantity, thus yielding the scalar-tensor versions of the torsion-based teleparallel gravity \cite{Geng:2011aj,Hohmann:2018rwf} and nonmetricity-based symmetric teleparallel gravity \cite{Jarv:2018bgs}. Like $f(\lc{R})$ gravity can be represented as a particular class of scalar-tensor theory, the  $f(T)$ and $f(Q)$ gravities also turn out to be particular classes of the corresponding teleparallel scalar-tensor theories. More elaborate extensions have been considered as well, and for a substantial review of models and applications one can turn to Ref.\ \cite{Bahamonde:2021gfp}. In general the field equations of these extended theories are different from their counterparts in the extensions of the original formulation of GR, and thus can accommodate brand new black hole solutions that await to be explored.

As the teleparallel versions of extended theories involve parts of the affine connection which are independent of the metric, there are two important aspects to stress. The first aspect is that the independent connection comes with its own field equations which must be properly solved to obtain a valid solution (in the local frame language in torsion teleparallelism the spin connection equations coincide with the antisymmetric tetrad equations \cite{Golovnev:2017dox,Hohmann:2017duq,Hohmann:2018rwf}, but the point is the same). Several authors have been led astray by a temptation to assume a simple diagonal metric or tetrad along with a vanishing connection, but these two together seldom actually satisfy all the equations. Another common situation that often occurs is that while seeking to satisfy the connection equations the remaining field equations get reduced to those equivalent of GR, and thus the configuration reproduces an already known solution. This is not a mistake in itself, but such results are somewhat trivial, since the extended theories typically contain GR as a particular limit, and thus naturally harbour all GR solutions in their fold.

The second aspect is that the independent connection may in principle possess different symmetry properties than the metric \cite{Hohmann:2019nat}. Although the possibility of solutions with mismatching symmetry may be entertained, a useful guiding principle in finding correct and nontrivial solutions of the field equations has been to assume that the connection obeys the same symmetry as the metric. In these lines the most general spherically symmetric and static connection compatible with vanishing curvature and vanishing nonmetricity, as written down in Refs.\ \cite{Ferraro:2011ks,Tamanini:2012hg,Hohmann:2019nat,Hohmann:2019fvf}, has yielded nontrivial exact black hole solutions in torsion based $f(T)$ \cite{Bahamonde:2021srr} and scalar-tensor gravities \cite{Bahamonde:2022lvh}, as well as provided an appropriate starting point for perturbative weak field explorations \cite{Ruggiero:2015oka, DeBenedictis:2016aze,Bahamonde:2019zea,Bahamonde:2020bbc,Golovnev:2021htv,Ren:2021uqb,Pfeifer:2021njm,DeBenedictis:2022sja,Zhao:2022gxl}. Recently the authors of Ref.\ \cite{DAmbrosio:2021zpm} derived the most general spherically symmetric and static connection compatible with vanishing curvature and vanishing torsion, and were able to report a class of exact nontrivial black hole solutions in nonmetricity based $f(Q)=Q^\kappa$ gravity, along with perturbative hairy corrections to the Schwarzschild solution in other $f(Q)$ models.

In this paper we focus upon the scalar-tensor version of symmetric teleparallel gravity where the curvature and torsion vanish, and assuming the metric, affine connection, as well as the scalar field all obey the properties of spherical symmetry and staticity, seek to find nontrivial black hole solutions that do not reduce to the ones already known in the context of general relativity. For this task, we also formulate no-hair theorems to obtain necessary conditions on the nonminimal couplings provided by the scalar field and on the form of its potential, whereby all the possible scalarised symmetric teleparallel solutions uncovered by the theorem must be asymptotically Minkowski.

We begin in Sec.~\ref{sec:symmetricTG} introducing the main geometrical notions of symmetric teleparallel gravity and its modification in the presence of a nonminimally coupled scalar field. We assume the metric tensor, affine connection and scalar field as the main independent quantities of the model, deriving the corresponding vacuum field equations to be solved in order to find new scalar hairy black holes, and pointing out the conditions to recover the $f(Q)$ model from this approach. In Sec.~\ref{sec:spherical symmetry} we consider the notion of symmetry for general metric-affine geometries, which allows the mathematical complexity of the field equations to be simplified for especially symmetric configurations. In particular, we restrict the symmetry to the case of static and spherically symmetric space-times and scalar fields, pointing out that the consideration of a flat and torsion-free affine connection turns out to provide two different sets of solutions. We derive a no-hair theorem for the model in Sec.~\ref{sec:hair} and consider it as a no-go result to constrain the search of asymptotically Minkowski scalarised black hole configurations. In section~\ref{sec:examination of field equations} we show that a preliminary examination of the field equations allows us to rule out the first set of static and spherically symmetric teleparallel connections for the obtainment of nontrivial solutions, whereas the second set is split into two branches, in such a way that only the second branch can provide asymptotically Minkowski solutions in agreement with the conclusions of the no-hair theorem. Accordingly, a thorough study of the field equations for this branch is addressed in Sec.~\ref{sec:scalar}. There are different ways to fully solve these equations, and by treating them case by case we are able to find different classes of exact solutions, such as the so-called Bocharova-Bronnikov-Melnikov-Bekenstein (BBMB) solution~\cite{bocharova1970exact,Bekenstein:1974sf,Bekenstein:1975ts}, which was first discovered in Riemannian conformal scalar-tensor theory and very recently also in torsional teleparallel scalar-tensor theory \cite{Bahamonde:2022lvh}. We also report several new black hole solutions which are characteristic of the present nonmetricity-scalar theory. Sec.~\ref{sec:theorem} is devoted to the demonstration of a theorem for $f(Q)$ gravity in spherical symmetry which constrains the metric to satisfy $g_{rr}\neq 1/g_{tt}$ to obtain solutions beyond GR. Finally, we end with summarizing conclusions in Sec.~\ref{sec:conclusions}.

We work in units $c=1$ with the metric signature $(-,+,+,+)$. Quantities denoted with a circle on top $\circ$ denote that they are defined with respect to the Levi-Civita connection, whereas quantities with a tilde on top are defined with respect to a general affine connection. On the other hand, quantities with no circle nor tilde are referred to symmetric teleparallel geometries (i.e. vanishing curvature and torsion tensors).

\section{symmetric teleparallel gravity and scalar-tensor theories}\label{sec:symmetricTG}

In the standard framework of GR the curvature tensor constructed from the metric-compatible and torsion-free connection turns out to be the geometrical quantity describing the gravitational field. On the other hand, symmetric teleparallel gravity assumes a flat nonmetric-compatible and torsion-free connection, which sets the nonmetricity tensor as the main geometrical quantity playing the role of representing the interaction:
\begin{equation}
    Q_{\rho\mu\nu} \equiv \nabla_{\rho} g_{\mu\nu} = \partial_{\rho}g_{\mu\nu}-\Gamma^{\lambda}\,_{\rho\mu}g_{\lambda\nu}-\Gamma^{\lambda}\,_{\rho\nu}g_{\mu\lambda}\,.
\end{equation}
Such a symmetric teleparallel connection is therefore characterised by vanishing curvature and torsion tensors
\begin{eqnarray}
    R^{\sigma}\,_{\rho\mu\nu} &\equiv& \partial_{\mu}\Gamma^{\sigma}\,_{\nu\rho}-\partial_{\nu}\Gamma^{\sigma}\,_{\mu\rho}+\Gamma^{\sigma}\,_{\mu\lambda}\Gamma^{\lambda}\,_{\nu\rho}-\Gamma^{\sigma}\,_{\mu\lambda}\Gamma^{\lambda}\,_{\nu\rho}=0\,,\\
    T^{\sigma}\,_{\mu\nu} &\equiv& \Gamma^{\sigma}\,_{\mu\nu}-\Gamma^{\sigma}\,_{\nu\mu}=0\,.
\end{eqnarray}
These conditions allow then to ascribe the Riemann tensor present in GR to the nonmetricity tensor as follows:
\begin{equation}
    \lc{R}^{\sigma}\,_{\rho\mu\nu}=\lc{\nabla}_{\nu}L^{\sigma}\,_{\mu\rho}-\lc{\nabla}_{\mu}L^{\sigma}\,_{\nu\rho}+L^{\sigma}\,_{\nu\lambda}L^{\lambda}\,_{\mu\rho}-L^{\sigma}\,_{\mu\lambda}L^{\lambda}\,_{\nu\rho}\,,
\end{equation}
where
\begin{equation}
    L^{\sigma}\,_{\mu\nu}=\frac{1}{2}\left(Q^{\sigma}\,_{\mu\nu}-Q_{\mu}\,^{\sigma}\,_{\nu}-Q_{\nu}\,^{\sigma}\,_{\mu}\right)\,,
\end{equation}
is the so-called disformation tensor. In particular, the scalar curvature of the Einstein-Hilbert action can be expressed as the sum of a nonmetricity scalar and a Riemannian divergence acting over the two independent traces of the nonmetricity tensor, which acts as a boundary term in the field equations derived from the mentioned action: 
\begin{eqnarray}\label{R and Q}
\lc{R}=Q+\lc{\nabla}_{\mu}(\hat{Q}^\mu-Q^\mu):=Q+B_Q\,,
\end{eqnarray}
where
\begin{equation}\label{Qscalar}
    Q \equiv -\,\frac{1}{4}\,Q_{\lambda\mu\nu}Q^{\lambda\mu\nu}+\frac{1}{2}\,Q_{\lambda\mu\nu}Q^{\mu\nu\lambda}+\frac{1}{4}\,Q_{\mu}Q^{\mu}-\frac{1}{2}\,Q_{\mu}\hat{Q}^{\mu}\,, \quad Q_{\mu} \equiv Q_{\mu\nu}\,^{\nu}\,, \quad \hat{Q}_{\mu} \equiv Q_{\nu\mu}\,^{\nu}\,,
\end{equation}
and we have defined the nonmetricity boundary term as $B_Q=\lc{\nabla}_{\mu}(\hat{Q}^\mu-Q^\mu)$. 

Accordingly, the framework of STEGR constructed from the nonmetricity scalar of a symmetric teleparallel connection provides the same field equations of GR and an equivalent description of the gravitational field:
\begin{equation}\label{STEGR_action}
    S = \frac{1}{2\kappa^2} \int\mathrm{d}^4x \sqrt{-g}\, Q + S_{\mathrm{m}}\,,
\end{equation}
where $\kappa^2=8\pi G$ and $S_{\mathrm{m}}$ denotes the action of matter. Here we assume that the matter action stays the same as in GR, i.e.\ it depends on the metric alone. Since the action (\ref{STEGR_action}) gives rise then to the same dynamics as GR, one can modify it in different ways to construct modified symmetric teleparallel theories of gravity.

Following these lines, an interesting extension arises when considering a scalar field nonminimally coupled to the nonmetricity scalar in the symmetric teleparallel framework, which accordingly leads to a scalar-nonmetricity theory whose field equations do not generally coincide with their counterparts provided by the curvature and torsion tensors~\cite{Jarv:2018bgs}:
\begin{equation}
\label{Action}
S = \frac{1}{2\kappa^2} \int\mathrm{d}^4x \sqrt{-g} \left( \mathcal A(\Phi) Q -\mathcal B(\Phi) g^{\alpha\beta}\partial_\alpha\Phi \partial_\beta\Phi 
-2{\mathcal V}(\Phi)\right) + S_{\mathrm{m}}\,.
\end{equation}
Like in the usual Riemannian scalar-tensor theory the nonminimal coupling function $\mathcal{A}$ sets the strength of the effective gravitational constant, $\mathcal{B}$ is the kinetic coupling function, and $\mathcal{V}$ is the scalar potential.
Specifically, the first field equation can be directly obtained from the variation of the action (\ref{Action}) with respect to the metric tensor, namely
\begin{eqnarray}
\nonumber
\kappa^2\mathcal{T}_{\mu\nu} &=&
\frac{2}{\sqrt{-g}}\nabla_{\alpha} \left( \sqrt{-g}\mathcal{A} P^{\alpha}_{\hphantom{\lambda}\mu\nu} \right)  - \frac{1}{2} g_{\mu\nu} \mathcal A Q  +  \mathcal{A} \left( P_{\mu\alpha\beta} Q_{\nu}^{\hphantom{\nu} \alpha\beta} - 2Q_{\alpha\beta\mu} P^{\alpha\beta}_{\hphantom{\alpha\beta} \nu} \right)  + \frac{1}{2} g_{\mu\nu} \left( \mathcal B g^{\alpha\beta}\partial_\alpha\Phi \partial_{\beta} \Phi + 2 {\mathcal V}  \right) - \mathcal{B} \partial_{\mu} \Phi \partial_\nu \Phi \,,\\
 \label{MetricFieldEq}
\end{eqnarray}
where we have defined the so-called superpotential (or conjugate) tensor as
\begin{eqnarray}
 P^\alpha{}_{\mu\nu}=-\,\frac{1}{4}Q^{\alpha}{}_{\mu\nu}+\frac{1}{2}Q_{(\mu}{}^\alpha{}_{\nu)}+\frac{1}{4}g_{\mu\nu}Q^\alpha-\frac{1}{4}(g_{\mu\nu}\hat{Q}^\alpha+\delta^\alpha{}_{(\mu}Q_{\nu)})\,.
\end{eqnarray}
By considering the identity
\begin{equation}
    \nabla_{\alpha}\sqrt{-g}=\frac{1}{2}\,Q_{\alpha}\sqrt{-g}\,,
\end{equation} 
we can rewrite the first term of the right-hand side of the field equation as
\begin{align}
    \frac{2}{\sqrt{-g}}\nabla_{\alpha} \left( \sqrt{-g}\mathcal{A}(\Phi) P^{\alpha}_{\hphantom{\lambda}\mu\nu} \right)=\mathcal{A}(\Phi)Q_{\lambda}P^\lambda{}_{\mu\nu}+2\bigl(\mathcal{A}(\Phi)\nabla_\lambda P^\lambda{}_{\mu\nu}+\frac{\dd\mathcal{A}}{\dd\Phi} P^\lambda{}_{\mu\nu} \partial_\lambda \Phi\bigr)\,.
\end{align}
Furthermore, we can introduce the Einstein tensor constructed from the metric tensor using the flatness and torsionless conditions as 
\begin{align}\label{einstein}
    \lc{G}_{\mu\nu}=2\nabla_\lambda P^\lambda{}_{\mu\nu}-\frac{1}{2}Qg_{\mu\nu}+P_{\rho\mu\nu}Q^{\rho\sigma}{}_{\sigma}+P_{\nu\rho\sigma}Q_\mu{}^{\rho\sigma}-2P_{\rho\sigma\mu}Q^{\rho\sigma}{}_{\nu} 
\end{align}
These results yield that the first field equation acquires the following compact form:
\begin{eqnarray}
\kappa^2\mathcal{T}_{\mu\nu} &=&\mathcal{A}(\Phi)\lc{G}_{\mu\nu}
+2\frac{\dd\mathcal{A}}{\dd\Phi} P^\lambda{}_{\mu\nu} \partial_\lambda \Phi + \frac{1}{2} g_{\mu\nu} \left( \mathcal B(\Phi) g^{\alpha\beta}\partial_\alpha\Phi \partial_{\beta} \Phi + 2 {\mathcal V}(\Phi)  \right) - \mathcal{B}(\Phi) \partial_{\mu} \Phi \partial_\nu \Phi \,.
 \label{MetricFieldEqF}
\end{eqnarray}
We see that the function $\mathcal{A}(\Phi)$ really gives dynamics to the effective gravitational constant, i.e.\ it measures how strongly does the matter energy momentum $\mathcal{T}_{\mu\nu}$ affect the Levi-Civita Einstein tensor $\lc{G}_{\mu\nu}$. At the zeroes and singularities of $\mathcal{A}$ the gravity as we know it ceases to exist, while negative $\mathcal{A}$ implies a sort of ``antigravity". When the scalar field is globally constant, the Eq.\ \eqref{MetricFieldEqF} reduces to the Einstein's equation in general relativity with the value of the potential playing the role of the cosmological constant. Therefore, the solutions of GR are trivially also solutions of these scalar-tensor theories when the scalar field is a constant. Next, by taking variations with respect to the independent affine connection, it is possible to find the following field equation~\cite{Jarv:2018bgs}:
\begin{eqnarray}\label{connectionEq}
  \nabla_\alpha \nabla_\beta (\sqrt{-g}\mathcal{A}P^{\alpha\beta}{}_\mu)=  \nabla_\beta \left\{\partial_\alpha \mathcal{A}\left[\nabla_\mu \left( \sqrt{-g}g^{\alpha \beta} \right) -\delta^\alpha_{\ \mu} \nabla_{\gamma}\left( \sqrt{-g}g^{\gamma \beta} \right) \right] \right\}=0\,,
\end{eqnarray}
which can be rewritten as
\begin{equation}    \frac{1}{2}Q_\beta K_\mu{}^\beta+ \nabla_\beta K_\mu{}^\beta=0\,,
\end{equation}
where we have defined the tensor
\begin{align}
    K_\mu{}^{\beta} &=\Big[\frac{1}{2}Q_\mu g^{\alpha\beta}-\frac{1}{2}\delta^\alpha_\mu Q^\beta-Q_{\mu}\,^{\alpha\beta}+\delta^\alpha_\mu Q_{\gamma}\,^{\gamma\beta}\Big](\partial_\alpha \mathcal{A})
\end{align}
and used $ Q_{\lambda}\,^{\mu\nu}=-\,\nabla_{\lambda}g^{\mu\nu}$. Finally, if we take variations with respect to the scalar field we find
\begin{eqnarray}
\label{ScalarFieldEq}
2\mathcal B \lc{\square}\Phi 
+\frac{\dd\mathcal{B}}{\dd\Phi}g^{\alpha\beta}\partial_{\alpha}\Phi\partial_{\beta}\Phi + \frac{\dd\mathcal{A}}{\dd\Phi} Q -
2\frac{\dd\mathcal{V}}{\dd\Phi} =0 \,,
\end{eqnarray}
with $\lc{\square} \equiv \lc{\nabla}_{\alpha}\lc{\nabla}^{\alpha}$. 

Let us remark that we can recover the $f(Q)$ gravity from our scalar-tensor theory by setting~\cite{Jarv:2018bgs}
\begin{align}\label{transftofQ}
 \mathcal{A}=f_Q\,,\quad
 \mathcal{B}=0\,,\quad
 \mathcal{V}=\tfrac{1}{2} \left(Q f_Q-f\right)\,,\quad \Phi=Q\,,
\end{align}
which gives us the $f(Q)$ field equations expressed by
\begin{eqnarray}
\kappa^2\mathcal{T}_{\mu\nu} &=&f_Q\lc{G}_{\mu\nu}
+2f_{QQ} P^\lambda{}_{\mu\nu} \partial_\lambda Q + \frac{1}{2} g_{\mu\nu}\left(Q f_Q-f\right)\,,
 \label{MetricFieldEqf(Q)}\\
 0&=&\nabla_\mu \nabla_\nu \left(\sqrt{-g}f_Q P^{\mu\nu}{}_\alpha\right)\,,\label{ConnectionFieldEqf(Q)}
\end{eqnarray}
where $f_Q=df/dQ$ and the scalar field equation \eqref{ScalarFieldEq} is identically satisfied. From now on we will study the vacuum case with $\mathcal{T}_{\mu\nu}=0$ (i.e\ no other fields besides $\Phi$).

\section{Spherical symmetry in symmetric teleparallel gravity}\label{sec:spherical symmetry}

\subsection{The notion of symmetry}

In virtue of the highly nonlinear character of the field equations derived in the context of metric-affine geometries, a guiding principle is to impose certain symmetry conditions in order to find nontrivial solutions. In this regard, the class of space-time symmetries provided by the action of a Lie group on a differentiable manifold in the framework of symmetric teleparallelism is based on the invariance of the corresponding Cartan geometry modeled by principal bundle automorphisms~\cite{Hohmann:2015pva}. In particular, infinitesimal symmetries can be described by the invariance of the metric tensor $g_{\mu\nu}$ and of a general affine connection $\tilde{\Gamma}^{\lambda}\,_{\mu\nu}$ under the flow of certain sets of vector fields. In this regard, it is worthwhile to stress that these quantities do not need to obey symmetry conditions provided by a common set of vector fields, but they can be preserved under the action of different groups of symmetries (e.g. see~\cite{Hehl:1999sb,Bahamonde:2021qjk,Bahamonde:2022meb} for exact stationary and axially symmetric solutions, which in the limit of static and spherically symmetric metric carry magnetic dilation charges that do not preserve this symmetry for the nonmetricity tensor). In any case, here we consider the simplest assumption which establishes the invariance under the same set of vector fields $Z_\zeta$, with $\zeta=1,..,m$, which is realised by vanishing their Lie derivatives in the direction of $Z_\zeta$,
\begin{align}\label{LieD_mag}
\mathcal{L}_{Z_\zeta}g_{\mu\nu}=0\,,\qquad
\mathcal{L}_{Z_\zeta}\tilde{\Gamma}^{\lambda}\,_{\mu\nu}=0\,.
\end{align}
Indeed, the first symmetry condition involving the metric tensor implies that the Levi-Civita part of the affine connection is straightforwardly preserved by a group of isometries, in virtue of the vanishing of its Lie derivative~\cite{Yano1972notes}:
\begin{equation}
\mathcal{L}_{Z_\zeta}\lc{\Gamma}^{\lambda}\,_{\mu\nu}=\frac{1}{2}g^{\lambda\rho}\left(\lc{\nabla}_{\mu}\mathcal{L}_{Z_\zeta}g_{\rho\nu}+\lc{\nabla}_{\nu}\mathcal{L}_{Z_\zeta}g_{\rho\mu}-\lc{\nabla}_{\rho}\mathcal{L}_{Z_\zeta}g_{\mu\nu}\right)\,,
\end{equation}
This means that the post-Riemannian part of the affine connection allows one to consider an additional symmetry condition for the connection, as is shown in the expression~\eqref{LieD_mag}.

Finally, a scalar field coupled to these geometrical quantities can also fulfil its own symmetry conditions. In particular, the corresponding invariance under the flow of the same set of vector fields $Z_\zeta$ is given by vanishing the Lie derivative\footnote{See \cite{Graham:2014ina} and \cite{Takahashi:2020hso} for other black hole solutions with scalar fields which do not fully satisfy the same symmetries of the metric tensor.}
\begin{equation}\label{LieD_sc}
\mathcal{L}_{Z_\zeta}\Phi=0\,,
\end{equation}
which in turn ensures the same invariance for its kinetic term $g^{\alpha\beta}\partial_\alpha\Phi\,\partial_\beta\Phi$ present in the action.

The fulfilment of the symmetry conditions (\ref{LieD_mag}) and (\ref{LieD_sc}) under the action of a particular Lie group constrains the form of the metric, nonmetricity and scalar fields displayed by the action (\ref{Action}), which turns out to be a key point in order to find different classes of scalarised solutions in the framework of symmetric teleparallel gravity.

\subsection{Two sets of static spherically symmetric flat and torsion-free affine connections}

The consideration of a spherically symmetric metric-affine geometry is realised by the invariance under the group $SO(3)$ of spatial rotations. In particular, by assuming spherical coordinates $(t,r,\theta,\phi)$ the corresponding infinitesimal spatial rotations are described by the Killing vectors forming the Lie algebra of the group, namely
\begin{align}
    R^\mu &= \left[ \begin{matrix} 0, & 0, & 0, & 1 \end{matrix} \right] \,, \\ 
    S^\mu &= \left[ \begin{matrix} 0, & 0, & \cos\phi, & -\sin\phi \cot\theta \end{matrix} \right]  \,, \\
    T^\mu &= \left[ \begin{matrix} 0, & 0, & -\sin\phi, & -\cos\phi \cot\theta \end{matrix} \right] \,.
\end{align}

The additional assumption of staticity is established by the time Killing vector
\begin{align}
\label{eq: time Killing vector}
    K^\mu &= \left[ \begin{matrix} 1, & 0, & 0, & 0 \end{matrix} \right]\,,
\end{align}
which  trivially sets the main geometrical quantities of the differentiable manifold to be independent of time. It is straightforward then to check that the symmetry conditions (\ref{LieD_mag}) and (\ref{LieD_sc}) referred to these Killing vectors allow the form of the metric tensor and the scalar field to be fixed, without any loss of generality, as
\begin{equation}
    ds^2=-\,g_{tt}(r)\dd t^2+g_{rr}(r) \dd r^2+r^2\dd \Omega^2\,, \qquad  
    \Phi=\Phi(r)\,, 
\end{equation}
where $\dd \Omega^2=\dd\theta^2+\sin^2\theta \,\dd\phi^2$ and $0\leq\theta\leq\pi$, $0\leq\phi\leq 2\pi$. Thus, the independent components of the metric tensor are described by two arbitrary functions depending on the radial coordinate, which also holds for the additional function describing the scalar field. Likewise, a direct computation of the Eq.~(\ref{LieD_mag}) without the imposition of the flatness and torsionless conditions over the affine connection gives rise to twenty arbitrary independent functions $\{\mathcal{C}_i(r)\}_{i=1}^{20}$ for this quantity~\cite{Hohmann:2019fvf}:
\begin{align}
\tilde{\Gamma}{}^{\lambda}{}_{\mu\nu} &=\left[\begin{matrix}
   \left[\begin{matrix}\mathcal{C}_1 & \mathcal{C}_2 & 0 & 0\\\mathcal{C}_3 & \mathcal{C}_4 & 0 & 0\\0 & 0 & \mathcal{C}_9 & -\mathcal{C}_{19} \sin \theta\\0 & 0 & \mathcal{C}_{19} \sin \theta & \mathcal{C}_{9} \sin^2 \theta \end{matrix}\right] & 
   \left[\begin{matrix} \mathcal{C}_5 & \mathcal{C}_6 & 0 & 0\\\mathcal{C}_7 & \mathcal{C}_8 & 0 & 0\\0 & 0 & \mathcal{C}_{10} & -\mathcal{C}_{20} \sin \theta\\0 & 0 & \mathcal{C}_{20} \sin \theta & \mathcal{C}_{10} \sin^2 \theta \end{matrix}\right] & 
    \end{matrix}\right. \nonumber \\
&  \qquad \left.\begin{matrix} & 
   \left[\begin{matrix}0 & 0 & \mathcal{C}_{11} & -\mathcal{C}_{15}\sin\theta\\0 & 0 & \mathcal{C}_{12} & -\mathcal{C}_{16}\sin\theta\\ \mathcal{C}_{13} & \mathcal{C}_{14} & 0 & 0\\ -\mathcal{C}_{17}\sin\theta & -\mathcal{C}_{18}\sin\theta & 0 & - \sin{\theta} \cos{\theta}\end{matrix}\right] &
   \left[\begin{matrix}0 & 0 & \vspace{0.1cm}\frac{\mathcal{C}_{15}}{\sin\theta} & \mathcal{C}_{11}\\\vspace{0.1cm} 0 & 0 & \frac{\mathcal{C}_{16}}{\sin\theta} & \mathcal{C}_{12}\\ \vspace{0.1cm}\frac{\mathcal{C}_{17}}{\sin\theta} & \frac{\mathcal{C}_{18}}{\sin\theta} & 0 & \cot{\theta}\\\mathcal{C}_{13} & \mathcal{C}_{14} & \cot{\theta} & 0\end{matrix}\right]\end{matrix}\right]
   \label{eq: Gamma set10}\,,
\end{align}
where the four matrices of Eq~\eqref{eq: Gamma set10} give the components of the affine connection in the order $\tilde{\Gamma}^{t}{}_{\mu\nu}$, $\tilde{\Gamma}^{r}{}_{\mu\nu}$, $\tilde{\Gamma}^{\theta}{}_{\mu\nu}$ and $\tilde{\Gamma}^{\phi}{}_{\mu\nu}$, respectively. Thereby, the corresponding spherically symmetric teleparallel connection $\Gamma^{\lambda}\,_{\mu\nu}$ must additionally vanish the torsion and curvature tensors, which restricts the form of expression~(\ref{eq: Gamma set10}) for this connection. Furthermore, given the fact that the curvature tensor contains quadratic terms in the affine connection, the last constraint admits two independent sets of solutions~\cite{DAmbrosio:2021zpm}\footnote{Note that the existence of different sets of symmetric solutions is indeed a common feature in other teleparallel frameworks~\cite{Hohmann:2019nat}.}. From a physical point of view, it is worthwhile to stress then that these two sets shall generally provide different physical configurations when imposing the corresponding field equations \eqref{MetricFieldEqF}, \eqref{connectionEq} and \eqref{ScalarFieldEq}. In any case, it is essential to note that in the presence of a nonminimally coupled scalar field the Birkhoff's theorem does not hold for the present model, which enables the existence of nontrivial static and spherically symmetric vacuum solutions beyond the Schwarzschild geometry of STEGR.

\subsubsection{Connection set 1}\label{sec:set1}

The first static and spherically symmetric set of solutions with vanishing torsion and curvature tensors can be parameterised by three independent components $\Gamma^{\phi}{}_{r \phi}$, $\Gamma^{t}{}_{rr}$ and $\Gamma^{r}{}_{rr}$, which are functions of the radial coordinate~\cite{DAmbrosio:2021zpm}:
\begin{align}
\Gamma^{\lambda}{}_{\mu\nu} &=
   \left[\begin{matrix}\left[\begin{matrix}c & \Gamma^{\phi}{}_{r \phi} & 0 & 0\\\Gamma^{\phi}{}_{r \phi} & \Gamma^{t}{}_{rr} & 0 & 0\\0 & 0 & - \frac{1}{c} & 0\\0 & 0 & 0 & - \frac{\sin^{2}{\theta}}{c}\end{matrix}\right] & \left[\begin{matrix}0 & 0 & 0 & 0\\0 & \Gamma^{r}{}_{rr} & 0 & 0\\0 & 0 & 0 & 0\\0 & 0 & 0 & 0\end{matrix}\right] & \left[\begin{matrix}0 & 0 & c & 0\\0 & 0 & \Gamma^{\phi}{}_{r \phi} & 0\\c & \Gamma^{\phi}{}_{r \phi} & 0 & 0\\0 & 0 & 0 & - \sin{\theta} \cos{\theta}\end{matrix}\right] & \left[\begin{matrix}0 & 0 & 0 & c\\0 & 0 & 0 & \Gamma^{\phi}{}_{r \phi}\\0 & 0 & 0 & \cot{\theta}\\c & \Gamma^{\phi}{}_{r \phi} & \cot{\theta} & 0\end{matrix}\right]\end{matrix}\right]
   \label{eq: Gamma set1}\,,
\end{align}
provided the constraint
\begin{align}
    (\Gamma^{\phi}{}_{r \phi})' &= c \Gamma^{t}{}_{rr} - \Gamma^{\phi}{}_{r \phi} ( \Gamma^{\phi}{}_{r \phi} - \Gamma^{r}{}_{rr}) \,,
    \label{eq: Gamma relations set1}
\end{align}
where the prime symbol denotes differentiation with respect to $r$. Using the definition~\eqref{Qscalar}, the nonmetricity scalar in this connection set 1 has the following form
\begin{equation}
    Q=\frac{ (4-3 r \Gamma^{\phi}{}_{r \phi})g_{tt}'}{2r g_{rr} g_{tt}}
    -\frac{ 1}{r^2 g_{rr}}\left(3 c r^2 \Gamma^{t}{}_{rr}-3 r^2 (\Gamma^{\phi}{}_{r \phi})^2+3 r (r \Gamma^{r}{}_{rr}+2) \Gamma^{\phi}{}_{r \phi}-2\right)+\frac{3 \Gamma^{\phi}{}_{r \phi}g_{rr}'}{2g_{rr}^2}+\frac{2}{r^2}\,,
\end{equation}
while the boundary term $B_Q$ becomes
\begin{eqnarray}
B_Q&=&\frac{1}{2 r^2 g_{rr}^2 g_{tt}^2}\Big[r g_{tt} g_{rr}' \left(r g_{tt}'+g_{tt} (4-3 r \Gamma^{\phi}{}_{r \phi})\right)+g_{rr} \Big(r^2 g_{tt}'^2+r g_{tt} \left((3 r \Gamma^{\phi}{}_{r \phi}-8) g_{tt}'-2 r g_{tt}''\right)\nonumber\\
&&+2 g_{tt}^2 \left(3 r^2 \Gamma^{\phi}{}_{r \phi}'+6 r \Gamma^{\phi}{}_{r \phi}-4\right)\Big)\Big]\,.
\end{eqnarray}
Clearly, by taking $Q+B_Q$ we recover the standard Ricci scalar computed in terms of the Levi-Civita connection $\lc{R}$:
\begin{eqnarray}
\lc{R}=\frac{r g_{tt} g_{rr}' \left(r g_{tt}'+4 g_{tt}\right)+g_{rr} \left(r^2 g_{tt}'^2-2 r g_{tt} \left(r g_{tt}''+2 g_{tt}'\right)-4 g_{tt}^2\right)+4 g_{rr}^2 g_{tt}^2}{2 r^2 g_{rr}^2 g_{tt}^2}\,,\label{ricci}
\end{eqnarray}
which does not depend on the connection components, as expected.

\subsubsection{Connection set 2}\label{sec:set2}
The second static and spherically symmetric set with vanishing curvature and torsion tensors is described by the following form of the affine connection~\cite{DAmbrosio:2021zpm}:
\begin{align}
\Gamma^{\lambda}{}_{\mu\nu} =&
\left[\left[\begin{matrix}- c \left(2 c - k\right) \Gamma^{t}{}_{\theta \theta}  - c + k & \frac{\left(2 c - k\right) \left(c \Gamma^{t}{}_{\theta \theta}  + 1\right) \Gamma^{t}{}_{\theta \theta} }{\Gamma^{r}{}_{\theta \theta} } & 0 & 0\\\frac{\left(2 c - k\right) \left(c \Gamma^{t}{}_{\theta \theta}  + 1\right) \Gamma^{t}{}_{\theta \theta} }{\Gamma^{r}{}_{\theta \theta} } & \Gamma^{t}{}_{rr}  & 0 & 0\\0 & 0 & \Gamma^{t}{}_{\theta \theta}  & 0\\0 & 0 & 0 & \Gamma^{t}{}_{\theta \theta}  \sin^{2}{\theta}\end{matrix}\right] \right. \nonumber \\ & \qquad
\left[\begin{matrix}- c \left(2 c - k\right) \Gamma^{r}{}_{\theta \theta}  & c \left(2 c - k\right) \Gamma^{t}{}_{\theta \theta}  + c & 0 & 0\\c \left(2 c - k\right) \Gamma^{t}{}_{\theta \theta}  + c & \Gamma^{r}{}_{rr}  & 0 & 0\\0 & 0 & \Gamma^{r}{}_{\theta \theta}  & 0\\0 & 0 & 0 & \Gamma^{r}{}_{\theta \theta}  \sin^{2}{\theta}\end{matrix}\right] \nonumber \\ & \qquad \left.
\left[\begin{matrix}0 & 0 & c & 0\\0 & 0 & \frac{- c \Gamma^{t}{}_{\theta \theta}  - 1}{\Gamma^{r}{}_{\theta \theta} } & 0\\c & \frac{- c \Gamma^{t}{}_{\theta \theta}  - 1}{\Gamma^{r}{}_{\theta \theta} } & 0 & 0\\0 & 0 & 0 & - \sin{\theta} \cos{\theta}\end{matrix}\right] \quad 
\left[\begin{matrix}0 & 0 & 0 & c\\0 & 0 & 0 & \frac{- c \Gamma^{t}{}_{\theta \theta}  - 1}{\Gamma^{r}{}_{\theta \theta} }\\0 & 0 & 0 & \cot{\theta}\\c & \frac{- c \Gamma^{t}{}_{\theta \theta}  - 1}{\Gamma^{r}{}_{\theta \theta} } & \cot{\theta} & 0\end{matrix}\right] \right]\,,
\label{eq: Gamma set2}
\end{align}
where $\Gamma^{t}{}_{\theta \theta}, \Gamma^{r}{}_{\theta\theta}, \Gamma^{r}{}_{rr}$ depend on $r$, and are related by\footnote{Note that the corresponding Eq.~\eqref{eq: Gamma relations set2B} derived in~\cite{DAmbrosio:2021zpm} contains a minor typo, since the respective factor $c$ multiplying the term $\left(2c-k\right)\Gamma^{t}{}_{\theta\theta}$ in the parenthesis is missing.}
\begin{subequations}
\begin{align}
    ( \Gamma^{r}{}_{\theta \theta})'  &= -\, c \left[\left(2 c - k\right) \Gamma^{t}{}_{\theta \theta}  + 2\right] \Gamma^{t}{}_{\theta \theta}  - \Gamma^{r}{}_{\theta \theta}  \Gamma^{r}{}_{rr}  - 1\,, \label{relation1}\\
    ( \Gamma^{t}{}_{\theta \theta})'  &= -\, \frac{\left[\left(c \left(2 c - k\right) \Gamma^{t}{}_{\theta \theta}  + 3 c - k\right) \Gamma^{t}{}_{\theta \theta}  + 1\right] \Gamma^{t}{}_{\theta \theta} }{\Gamma^{r}{}_{\theta \theta} } - \Gamma^{r}{}_{\theta \theta}  \Gamma^{t}{}_{rr} \,.\label{eq: Gamma relations set2B}
\end{align}
\label{eq: Gamma relations set2}
\end{subequations}
For this set, the nonmetricity scalar becomes
\begin{eqnarray}
Q&=&\frac{1}{4} \bigg\{\frac{1}{r^2 g_{rr} g_{tt} (\Gamma^{r}{}_{\theta \theta})^2}\Big[4 g_{tt} \Big(r (\Gamma^{t}{}_{\theta \theta})^2 \left(r \left(6 c^2-4 c k+k^2\right)-c (2 c-k) (r \Gamma^{r}{}_{rr}+2) \Gamma^{r}{}_{\theta \theta}\right)+c^2 r^2 (k-2 c)^2 (\Gamma^{t}{}_{\theta \theta})^4\nonumber\\
&&+2 c r^2 (k-2 c)^2 (\Gamma^{t}{}_{\theta \theta})^3+r \Gamma^{t}{}_{\theta \theta} (\Gamma^{r}{}_{\theta \theta} (-2 c r (k-2 c) \Gamma^{t}{}_{rr} \Gamma^{r}{}_{\theta \theta}+k r \Gamma^{r}{}_{rr}+2 k)+4 c r)\nonumber\\&&+\Gamma^{r}{}_{\theta \theta} \Big(\Gamma^{r}{}_{\theta \theta} \left(\Gamma^{r}{}_{\theta \theta}g_{rr}'-k r^2 \Gamma^{t}{}_{rr}+2\right)
+2 r^2 \Gamma^{r}{}_{rr}+4 r\Big)+2 r^2\Big)+2 r \Gamma^{r}{}_{\theta \theta} (c r (2 c-k) (\Gamma^{r}{}_{\theta \theta})^2g_{rr}'\nonumber\\
&&+g_{tt}' (r (c \Gamma^{t}{}_{\theta \theta}+1) ((k-2 c) \Gamma^{t}{}_{\theta \theta}+2)+4 \Gamma^{r}{}_{\theta \theta}))\Big]-\frac{2 (c \Gamma^{t}{}_{\theta \theta}+1)g_{rr}' ((k-2 c) \Gamma^{t}{}_{\theta \theta}+2)}{g_{rr}^2 \Gamma^{r}{}_{\theta \theta}}\nonumber\\
&&+\frac{4 \Gamma^{r}{}_{\theta \theta}g_{tt}'-4 c r (2 c-k) (c r \Gamma^{t}{}_{\theta \theta} (2 c \Gamma^{t}{}_{\theta \theta}-k \Gamma^{t}{}_{\theta \theta}+2)+(r \Gamma^{r}{}_{rr}-2) \Gamma^{r}{}_{\theta \theta}+r)}{r^2 g_{tt}}+\frac{2 c (k-2 c) \Gamma^{r}{}_{\theta \theta}g_{tt}'}{g_{tt}^2}\nonumber\\
&&+\frac{8 c \Gamma^{t}{}_{\theta \theta} ((k-2 c) \Gamma^{t}{}_{\theta \theta}-2)-8 \Gamma^{r}{}_{rr} \Gamma^{r}{}_{\theta \theta}}{r^2}\bigg\}\,,\label{Q2}
\end{eqnarray}
while the boundary term $B_Q$ behaves as
\begin{eqnarray}
B_Q&=&-\frac{1}{2 r^2 g_{rr}^2 g_{tt}^2(\Gamma^{r}{}_{\theta \theta})^2}\Big[-r g_{tt}\Gamma^{r}{}_{\theta \theta} g_{rr}' \left(g_{tt} (r (c\Gamma^{t}{}_{\theta \theta}+1) ((k-2 c)\Gamma^{t}{}_{\theta \theta}+2)+4\Gamma^{r}{}_{\theta \theta})+r\Gamma^{r}{}_{\theta \theta} g_{tt}'\right)\nonumber\\
&&+g_{rr} \Big(r g_{tt}\Gamma^{r}{}_{\theta \theta} \left(c r (2 c-k)(\Gamma^{r}{}_{\theta \theta})^2 g_{rr}'+r (c\Gamma^{t}{}_{\theta \theta}+1) g_{tt}' ((k-2 c)\Gamma^{t}{}_{\theta \theta}+2)+2\Gamma^{r}{}_{\theta \theta} \left(r g_{tt}''+4 g_{tt}'\right)\right)\nonumber\\
&&+2 g_{tt}^2 \Big(r^2 (c\Gamma^{t}{}_{\theta \theta}+1)(\Gamma^{r}{}_{\theta \theta})' (2 c\Gamma^{t}{}_{\theta \theta}-k\Gamma^{t}{}_{\theta \theta}-2)+r\Gamma^{r}{}_{\theta \theta} \Big(r(\Gamma^{t}{}_{\theta \theta})' (2 c (k-2 c)\Gamma^{t}{}_{\theta \theta}+k)\nonumber\\
&&+2\Gamma^{t}{}_{\theta \theta} (c (k-2 c)\Gamma^{t}{}_{\theta \theta}+k)+4\Big)+(\Gamma^{r}{}_{\theta \theta})^3 g_{rr}'+4(\Gamma^{r}{}_{\theta \theta})^2\Big)-r^2(\Gamma^{r}{}_{\theta \theta})^2 g_{tt}'^2\Big)\nonumber\\
&&+g_{rr}^2(\Gamma^{r}{}_{\theta \theta})^2 \left(\Gamma^{r}{}_{\theta \theta} \left(g_{tt}' \left(c r^2 (k-2 c)+2 g_{tt}\right)+4 c r (2 c-k) g_{tt}\right)+2 g_{tt}(\Gamma^{r}{}_{\theta \theta})' \left(c r^2 (2 c-k)+2 g_{tt}\right)\right)\Big]\,.\label{BQ2}
\end{eqnarray}
Again, one can easily check that the sum of the expressions~\eqref{Q2} and \eqref{BQ2} gives rise to the Ricci scalar given by~\eqref{ricci}.

\section{No-hair theorem for symmetric teleparallel gravity}\label{sec:hair}
In this section we will provide a no-hair theorem for our theory that will constrain the functions $\mathcal{A}$, $\mathcal{B}$ and $\mathcal{V}$ for which it will be possible to find asymptotically Minkowski black hole solutions. Similarly to the procedure done in \cite{Bahamonde:2022lvh} (or \cite{Herdeiro:2015waa}) we will first find the trace of the vacuum field equations \eqref{MetricFieldEqF}
\begin{equation}
    \mathcal{A}\lc{G}^\mu{}_\mu+\frac{\dd\mathcal{A}}{\dd\Phi}(\partial_\mu \Phi) \left( Q^\mu - \hat{Q}^\mu \right) + \mathcal{B} \partial_\mu \Phi \partial^\mu \Phi + 4\mathcal{V} =0\,,
\end{equation}
which can be rewritten by using $\lc{G}^\mu{}_\mu=-\lc{R}$ and the relationship between the Ricci and nonmetricity scalars \eqref{R and Q}, yielding
\begin{equation} \label{trace of field equations}
    \mathcal{A}Q-\mathcal{B}\partial_\mu \Phi \partial^\mu \Phi - 4\mathcal{V} = \lc{\nabla}_\mu \left[ \mathcal{A}\left( Q^\mu - \hat{Q}^\mu \right) \right]\,.
\end{equation}
Next we will integrate this equation over the exterior region of the black hole which we will denote by $V$. Notice that we can use the divergence theorem to replace the right hand side of the expression with an integral over the boundary of $V$, denoted by $\partial V$. This procedure applied to the above equation gives us
\begin{equation} \label{trace of field equations integrated}
    \int_V \dd^4x \sqrt{-g}\left( \mathcal{A}Q-\mathcal{B}\partial_\mu \Phi \partial^\mu \Phi - 4\mathcal{V} \right) = \int_{\partial V}\mathcal{A}\left( Q^\mu - \hat{Q}^\mu \right)n_{\mu} \dd^3x\,,
\end{equation}
where $n_\mu$ denotes the components of the outward pointing normal vector of the boundary hypersurface. The boundary $\partial V$ can be taken to consist of the spatial asymptotics of $r \to \infty$, the event horizon of the black hole, plus the regions in the past and future connecting them. In fact, the contributions to the integral coming from the past and future hypersurfaces cancel out, since the timelike normal vector points in the opposite direction at the past and future, while the rest of the expression is the same due to the static property of the spacetime. In the present case the event horizon coincides with the Killing horizon, i.e.\ the null hypersurface where the norm of the Killing vector \eqref{eq: time Killing vector} vanishes. Therefore on the horizon part of the boundary hypersurface the normal vector $n^\mu$ is given by $K^\mu$ and has only one nonzero component, namely $n^t$. On the spatial asymptotics part of the boundary hypersurface the normal vector is pointing radially outwards, and thus has only $n^r$ as a nonzero component.
Thus in order to evaluate the right hand side of \eqref{trace of field equations integrated} one has to analyse the behaviour of the integrand at the event horizon and asymptotics. 

We can repeat the same process for the scalar-field equation \eqref{ScalarFieldEq} after multiplying it with $\Phi$ and rewriting this equation as 
\begin{equation}\label{scalar field equation different form}
\partial_\mu \Phi \partial^\mu \Phi \left(\frac{1}{2}\frac{\dd\mathcal{B}}{\dd\Phi}\Phi+\mathcal{B} \right)  - \frac{1}{2}\frac{\dd\mathcal{A}}{\dd\Phi}\Phi Q+\frac{\dd\mathcal{V}}{\dd\Phi}\Phi = \lc{\nabla}_\mu \left( \mathcal{B}\Phi \lc{\nabla}^\mu \Phi \right)\,.
\end{equation}
Integrating this equation over the exterior region $V$ and using the divergence theorem, yields
\begin{equation} \label{scalar field equation different form integrated}
    \int_V \dd^4x \sqrt{-g}\left[ \partial_\mu \Phi \partial^\mu \Phi \left(\frac{1}{2}\frac{\dd\mathcal{B}}{\dd\Phi}\Phi+\mathcal{B} \right)  - \frac{1}{2}\frac{\dd\mathcal{A}}{\dd\Phi}\Phi Q+\frac{\dd\mathcal{V}}{\dd\Phi}\Phi \right]= \int_{\partial V} \mathcal{B}(\Phi \lc{\nabla}^\mu \Phi) n_{\mu} \dd^3 x\,.
\end{equation}
In the current paper we are interested in asymptotically Minkowski black hole solutions and thus in the context of no-hair theorems we would also satisfy that assumption. Therefore one can see that the right hand side of equation \eqref{scalar field equation different form integrated} becomes zero at the asymptotics since the scalar field should become constant (otherwise the total energy of the configuration would diverge and the asymptotic Minkowski cannot be realised).
Since the spherically symmetric scalar field depends only on $r$ and the only relevant component of $n_\mu$ at the event horizon is $n_t$, the right hand side of~\eqref{scalar field equation different form integrated} does not get any contribution there either.

Eq.~\eqref{trace of field equations integrated} depends on the form of the combination $(Q^\mu - \hat{Q}^\mu)n_\mu$, and since we have two different sets of connections, we will have different expressions evaluated at the horizon and asymptotics. This will be studied further in Sec.~\ref{sec:scalar} where we will analyse what happens to the field equations for each set.

\section{Examination of field equations}\label{sec:examination of field equations}
In Sec.\ \ref{sec:spherical symmetry}, we stressed that in spherical symmetry there are two different sets of connections which are endowed with vanishing curvature and torsion, while in Sec.\ \ref{sec:hair} we derived two integral expressions which asymptotically Minkowski black hole solutions must satisfy. In this section, we examine the equations further in view of the different connections.

\subsection{Off-diagonal metric equation}\label{sec:off-diagonal equation}

For the connection set 1 (see Sec.~\ref{sec:set1}), the metric field equations underlying from the action~\eqref{Action} present the following off-diagonal $tr$ component:
\begin{align}
\label{eq: tr set 1}
    \frac{3}{2} \frac{\dd\mathcal{A}}{\dd\Phi}\Phi'(r)=0 \,.
\end{align}
It implies that either $\mathcal{A}(\Phi)=\textrm{const}$ and thus the nonminimal coupling has no effect on the configuration besides the overall redefinition of the gravitational constant, or the  scalar field is constant everywhere, $\Phi=\textrm{const}$, and again the only effect is the overall adjustment of the gravitational constant. Thus these two cases are trivial in the sense that effectively are reduced to GR with a minimally coupled scalar field. These configurations are therefore completely equivalent to the case of GR in the presence of a minimally coupled scalar field, and accordingly we will not pursue them any further here since they have been extensively studied in the literature~\cite{Sotiriou:2011dz,Herdeiro:2015waa}. In effect, the same was observed in the $f(Q)$ case, where the connection set 1 reduced the theory to GR \cite{DAmbrosio:2021zpm}.

On the other hand, the component $tr$ of the metric field equation~\eqref{MetricFieldEqF} for the connection set 2 (see Sec.~\ref{sec:set2}) turns out to acquire the following form:
\begin{align}
\label{eq: off-diagonal equation set 2}
  \frac{1}{2}  \left[2 c (k-2 c) \Gamma^{t}{}_{\theta\theta}+k\right] \frac{\dd\mathcal{A}}{\dd\Phi}\Phi'(r)=0\,.
\end{align}
This equation can be solved in different ways. The trivial way is given by the cases $\mathcal{A}(\Phi)=\textrm{const}$ or  $\Phi=\textrm{const}$, which again effectively reduces the model to GR with a minimally coupled scalar field, and will not be considered further here. The other way is to impose that the front factor in the above equation vanishes. Like in the $f(Q)$ theory setting \cite{DAmbrosio:2021zpm}, in this case, one may distinguish two different subcases or {\it branches}.

\subsection{The first branch}\label{sec:firstbranch}
By considering the connection set 2, the first nontrivial way to solve Eq.~\eqref{eq: off-diagonal equation set 2} is
\begin{align}
\label{eq: set 2 branch 1}
   \Gamma^{t}{}_{\theta\theta} &=\frac{k}{2 c (2 c-k)}\,,\quad 2c\neq k\,,\quad k\neq0\,.
\end{align}
The relations \eqref{eq: Gamma relations set2} can then be used to express
\begin{subequations}
\begin{align}
    \Gamma^{r}{}_{rr} &= - \frac{( 8 c^{2} - 4 c k ) (\Gamma^{r}{}_{\theta \theta})' + 8 c^{2} + k^{2}}{4 c \left(2 c - k\right) \Gamma^{r}{}_{\theta \theta}} \,, \\
   \Gamma^{t}{}_{rr} &= - \frac{k \left(8 c^{2} + 2 c k - k^{2}\right)}{8 c^{2} \left(2 c - k\right)^{2} \left(\Gamma^{r}{}_{\theta \theta}\right)^{2}} \,
\end{align}
\end{subequations}
and with such substitutions the difference in the nonmetricity vectors is
\begin{align}
    Q_\mu - \hat{Q}_\mu &= \left[\begin{matrix}\frac{k \left(2 c + k\right) \left(4 c - k\right) g_{tt}}{8 c^{2} \left(2 c - k\right)^{2} \left(\Gamma^{r}{}_{\theta \theta}\right)^{2} g_{rr}}- \frac{12 c^{3} r^{2} - 4 c^{2} k r^{2} - c k^{2} r^{2} + 2 k g_{tt}}{2 c r^{2} \left(2 c - k\right)}\,,\frac{r g_{tt}' + 4 g_{tt}}{r g_{tt}} + \frac{\left(2 c^{2} r^{2} - c k r^{2} + 2 g_{tt}\right) \Gamma^{r}{}_{\theta \theta} g_{rr}}{r^{2} g_{tt}} + \frac{\left(4 c - k\right)^{2}}{4 c \left(2 c - k\right) \Gamma^{r}{}_{\theta \theta}}\,,0\,, 0 \end{matrix}\right]\,.
\end{align}
We can use these expressions to examine Eq.\ \eqref{trace of field equations integrated}. All solutions of the full field equations do satisfy \eqref{trace of field equations integrated} (since that equation is derived from the trace of the metric field equation), but we are especially interested in the black hole solutions which have a horizon in the interior and reduce to Minkowski in the spatial asymptotics. In this case, they are easier to find, if the integrand on the RHS of \eqref{trace of field equations integrated} vanishes. Note that on the black hole horizon $g_{tt}=0$ and the only nonzero component of the normal vector $n^\mu$ is $n^t$. Thus we see that the integrand vanishes on the black hole horizon if $k=-6c$. On the other boundary where $r\to \infty$, the asymptotics provided by the Minkowski condition implies $g_{tt} \to 1$, $g_{rr} \to 1$. Then, it turns out that it is not possible to vanish the integrand with $k=-6c$ if the connection component $\Gamma^r{}_{\theta\theta}$ is real. The connection component $\Gamma^r{}_{\theta\theta}$ occurs frequently in the remaining field equations, which makes satisfying the equations and that condition a nontrivial task, if at all possible. For this reason we will not investigate the 1st branch of  connections any further here. The difficulty in dealing with the 1st branch equations was reported also in the $f(Q)$ case \cite{DAmbrosio:2021zpm}. 

\subsection{The second branch}\label{secondbranch}
The second nontrivial way to solve Eq.~\eqref{eq: off-diagonal equation set 2} is given by
\begin{align}
\label{eq: set 2 branch 2}
    c=k=0\,.
\end{align}
Now the conditions \eqref{eq: Gamma relations set2} can be used to express the corresponding components of the symmetric teleparallel connection as
\begin{subequations}
\label{eq: Gamma relations set 2 branch 2}
\begin{align}
    \Gamma^{r}{}_{rr} &= -\,\frac{(\Gamma^{r}{}_{\theta \theta})' + 1}{\Gamma^{r}{}_{\theta \theta}} \,, \\
   \Gamma^{t}{}_{rr} &= -\,\frac{\Gamma^{r}{}_{\theta \theta} (\Gamma^{t}{}_{\theta \theta})' + \Gamma^{t}{}_{\theta \theta}}{\left(\Gamma^{r}{}_{\theta \theta}\right)^{2}} \,.
\end{align}
\end{subequations}
With such substitutions the difference in the nonmetricity vectors is
\begin{align}
    Q_\mu - \hat{Q}_\mu &= \left[\begin{matrix}\frac{{g_{tt}} (\Gamma^{t}{}_{\theta \theta})'}{\Gamma^{r}{}_{\theta \theta} {g_{rr}}} + \frac{\Gamma^{t}{}_{\theta \theta} {g_{tt}}}{\left(\Gamma^{r}{}_{\theta \theta}\right)^{2} {g_{rr}}} - \frac{2 \Gamma^{t}{}_{\theta \theta} {g_{tt}}}{r^{2}}\,,  \frac{2}{\Gamma^{r}{}_{\theta \theta}} + \frac{r {g_{tt}'} + 4 {g_{tt}}}{r {g_{tt}}} + \frac{2 \Gamma^{r}{}_{\theta \theta} {g_{rr}}}{r^{2}}\,,  0\,,  0
    \end{matrix}\right]\,,
\end{align}
and the nonmetricity scalar~\eqref{Q2} simplifies as
\begin{align}
\label{eq: nonmetricity scalar set 2 branch 2}
    Q &=\frac{rg_{rr} g'_{tt} - r g_{tt} g'_{rr} + 4 g_{rr} g_{tt}}{r \Gamma^{r}{}_{\theta \theta} g_{rr}^{2} g_{tt}} - \frac{2 \left(r^{2} - \left(\Gamma^{r}{}_{\theta \theta}\right)^{2} g_{rr}\right) (\Gamma^{r}{}_{\theta \theta})'}{r^{2} \left(\Gamma^{r}{}_{\theta \theta}\right)^{2} g_{rr}} + \frac{\left(g_{rr} g'_{tt} + g_{tt} g'_{rr}\right) \Gamma^{r}{}_{\theta \theta}}{r^{2} g_{rr} g_{tt}}
    + \frac{2 \left(r g'_{tt} + g_{rr} g_{tt} + g_{tt}\right)}{r^{2} g_{rr} g_{tt}} \,.
\end{align}
Again, we can use these expressions to examine Eq.\ \eqref{trace of field equations integrated}. On the black hole horizon the integrand on the RHS of \eqref{trace of field equations integrated} automatically vanishes under the condition $g_{tt}=0$ and $n^\mu \sim n^t$. In the spatial asymptotics with the Minkowski condition $g_{tt} \to 1$, $g_{rr} \to 1$ we obtain that $Q^r - \hat{Q}^r$ vanishes if  $\Gamma^r{}_{\theta\theta}\to \pm r^q$ with $0<q<2$.
The connection components $\Gamma^r{}_{\theta\theta}$ appear in the field equations but we can reasonably expect that the equations allow configurations with such asymptotics, and thus are motivated to investigate this branch further. (The divergence of $\Gamma^r{}_{\theta\theta}$ in the asymptotics is almost expected, since for Minkowski metric in spherical coordinates the respective Levi-Civita connection component is $\lc\Gamma^r{}_{\theta\theta}=-r$.)

We have seen that for the connections defined by~\eqref{eq: Gamma set2}, \eqref{eq: Gamma relations set2}, \eqref{eq: set 2 branch 2} the boundary integrals on the RHS of \eqref{trace of field equations integrated} and \eqref{scalar field equation different form integrated} are zero. 
Therefore we can add the two equations and obtain 
\begin{equation}
\label{eq: no-hair integral}
    \int_V \dd^4x \sqrt{-g}\left[Q\left(\mathcal{A}-\frac{1}{2}\frac{\dd\mathcal{A}}{\dd\Phi}\Phi \right)+\frac{1}{2}\frac{\dd\mathcal{B}}{\dd\Phi}\Phi \partial_\mu \Phi \partial^\mu \Phi +\frac{\dd\mathcal{V}}{\dd\Phi}\Phi - 4\mathcal{V} \right]=0\,.
\end{equation}
This can be understood as a general necessary (but not sufficient) condition for a theory to have asymptotically Minkowski black hole solutions. 
If the above equation does not hold, the theory defined by the functions $\mathcal{A}$, $\mathcal{B}$, $\mathcal{V}$ will not have asymptotically Minkowski black hole solutions. A particularly promising option in expecting to find solutions is when the integral in \eqref{eq: no-hair integral} vanishes due to the identically zero integrand. This hunch becomes more clear when without the loss of generality we redefine the scalar field so that the kinetic coupling function becomes a constant, $\mathcal{B}=\beta$. The integral reduces to 
\begin{equation}
\label{eq: no-hair beta}
    \int_V \dd^4x \sqrt{-g}\left[Q\left( \mathcal{A}- \frac{1}{2}\frac{\dd\mathcal{A}}{\dd\Phi}\Phi \right) +\frac{\dd\mathcal{V}}{\dd\Phi}\Phi - 4 \mathcal{V} \right]=0\,.
\end{equation}
which immediately suggests that certain functions $\mathcal{A}(\Phi)$ and $\mathcal{V}(\Phi)$ can be especially ``friendly'' for yielding analytic black hole solutions. The nonmetricity scalar $Q$ given by \eqref{eq: nonmetricity scalar set 2 branch 2} is a complicated function of the metric and connection and will not typically vanish, but the first term in the integrand is still identically zero if
\begin{align}
    \mathcal{A}(\Phi)=C\Phi^2\,, 
\end{align}
where $C$ is some constant. Although in this work we were keen to try to solve the field equations with general $\mathcal{A}$, all the solutions we actually found and report in Sec.~\ref{secsol1} and \ref{secsol2}  were indeed the ones with quadratic coupling function. We can also see that fixing the coupling function to $\mathcal{A}\sim \Phi^2$ implies that the potential of the scalar field should better be of the form
\begin{align}
  \mathcal{V}(\Phi) &= \hat{C}\Phi^4 \,,
\end{align}
where $\hat{C}$ is some other constant. Due to the extent of the required effort we were not able to dig very deep in this direction, and thus only highlight theories with quartic potentials as potentially promising for future investigations.

\section{scalarised solutions for symmetric teleparallel gravity}\label{sec:scalar}
This section is devoted to find scalarised solutions in symmetric teleparallel gravity by considering the theory~\eqref{Action} for the spherically symmetric connection set 2 and branch 2 (see Secs.~\ref{sec:set2} and~\ref{secondbranch}).

\subsection{Field equations}
For this branch we find that the metric field equations~\eqref{MetricFieldEq} are
\begin{eqnarray}
    0&=&\,\frac{\mathcal{A}(\Phi)}{r^{2}g^{2}_{rr}}\left[
    rg'_{rr}+g_{rr}^2-g_{rr}\right]-\,\frac{\Phi'}{r^{2}g_{rr}\Gamma^{r}{}_{\theta \theta}}\left[r^{2}+2r\Gamma^{r}{}_{\theta \theta}+(\Gamma^{r}{}_{\theta \theta})^{2}g_{rr}\right]\frac{\dd\mathcal{A}}{\dd\Phi}-\frac{\mathcal{B}(\Phi)\Phi'^{2}}{2g_{rr}}-\mathcal{V}(\Phi)\,,\label{eqB1}\\
    0&=&\frac{\mathcal{A}(\Phi)}{r^{2}g_{tt}g_{rr}}\left[g_{tt}\left(g_{rr}-1\right)-rg'_{tt}\right]+\frac{\Phi'}{r^{2}g_{rr}\Gamma^{r}{}_{\theta \theta}}\left[r^2-g_{rr}(\Gamma^{r}{}_{\theta \theta})^{2}\right]\frac{\dd\mathcal{A}}{\dd\Phi}+\frac{\mathcal{B}(\Phi)\Phi'^{2}}{2g_{rr}}-\mathcal{V}(\Phi)\,,\label{eqB2}\\
    0&=&\frac{\mathcal{A}(\Phi)}{4rg^{2}_{tt}g^{2}_{rr}}\left[g_{tt}g'_{tt}\left(rg'_{rr}-2g_{rr}\right)+rg_{rr}g'^{\,2}_{tt}+2g^{2}_{tt}g'_{rr}-2rg_{tt}g_{rr}g''_{tt}\right]\nonumber\\
    &&-\,\frac{\Phi'}{2rg_{tt}g_{rr}\Gamma^{r}{}_{\theta \theta}}\left[rg'_{tt}\Gamma^{r}{}_{\theta \theta}+2g_{tt}\left(r+\Gamma^{r}{}_{\theta \theta}\right)\right]\frac{\dd\mathcal{A}}{\dd\Phi}-\frac{\mathcal{B}(\Phi)\Phi'^2}{2g_{rr}}-\mathcal{V}(\Phi)\,,\label{eqB3}
\end{eqnarray}
whereas the connection equation~\eqref{connectionEq} for this case becomes
\begin{align}
0=&2 \left(r^{2} - \left(\Gamma^{r}{}_{\theta \theta}\right)^{2}    {g_{rr}}  \right)  {g_{rr}}    {g_{tt}}   \left( \left(\Phi'  \right)^{2} \frac{d^{2}\mathcal{A}}{d \Phi^{2}}  + \Phi''  \frac{d \mathcal{A}}{d \Phi}  \right) \nonumber \\
&+ \left(r^{2}  {g_{rr}}   {g_{tt}'}   - r^{2}  {g_{tt}}   {g_{rr}'}   + 4 r  {g_{rr}}    {g_{tt}}   - \left(\Gamma^{r}{}_{\theta \theta}\right)^{2}    {g_{rr}}^{2}   {g_{tt}'}   - \left(\Gamma^{r}{}_{\theta \theta}\right)^{2}    {g_{rr}}    {g_{tt}}   {g_{rr}'}   - 4 \Gamma^{r}{}_{\theta \theta}    {g_{rr}}^{2}    {g_{tt}}   ( \Gamma^{r}{}_{\theta \theta})'  \right) \Phi'  \frac{d \mathcal{A}}{d \Phi} \,.\label{eqB4}
\end{align}
Here we have further used the condition~\eqref{relation1} for the connection component $\Gamma^{r}{}_{rr}$.

Furthermore, the scalar field equation~\eqref{ScalarFieldEq} reduces to
\begin{eqnarray}
    0=-\frac{\dd\mathcal{A}}{\dd\Phi}Q+\frac{\mathcal{B}(\Phi)\left\{r\Phi'g_{tt}g'_{rr}-g_{rr}\left[r\Phi'g'_{tt}+2g_{tt}\left(r\Phi''+2\Phi'\right)\right]\right\}}{rg_{tt}g^{2}_{rr}}-\frac{\Phi'^2}{g_{rr}}\frac{\dd\mathcal{B}}{\dd\Phi}+2\frac{\dd\mathcal{V}}{\dd\Phi}\,.\label{eqB5}
\end{eqnarray}
Among the five equations \eqref{eqB1}-\eqref{eqB5} only four are actually independent of each other. Given the theory functions $\mathcal{A}$, $\mathcal{B}$, $\mathcal{V}$ the equations should be solved for the four quantities $g_{tt}$, $g_{rr}$, $\Gamma^{r}{}_{\theta \theta}$, $\Phi$. The other connection components must obey the relations \eqref{eq: Gamma relations set 2 branch 2}, but in particular $\Gamma^{t}{}_{\theta \theta}$ remains completely undetermined by the equations. When discussing and presenting the solutions we will ignore these extra connection components from now on. However, the extra connection components appear in the boundary term \eqref{BQ2}, and thus arguments concerning black hole energy or entropy might still fix them (see e.g.\ Refs.\ \cite{BeltranJimenez:2021kpj,Gomes:2022vrc}).

The main problem while facing the equations \eqref{eqB1}-\eqref{eqB5} is that they combine the quantities in a complicated way, and it is very hard to disentangle the equations into a form where direct integration can be started. It is almost impossible to proceed without making some simplifying assumptions on the way, and thus sacrificing the generality of the solutions.
One approach to tackle the five equations \eqref{eqB1}-\eqref{eqB5} is that the connection component $\Gamma^{r}{}_{\theta \theta}$ can be solved in two different ways. The simplest solution is the one where one assumes that the connection component $\Gamma^{r}{}_{\theta\theta}$ does not depend on the function $\mathcal{A}(\Phi)$ that specifies  the theory. The solution for this case can be found by directly solving the connection equation~\eqref{eqB4} which reads surprisingly simply as
\begin{align}
     \Gamma^{r}{}_{\theta \theta} &= \pm \frac{r}{\sqrt{g_{rr}}}\,.\label{gamma}
\end{align}
Another approach is to sum \eqref{eqB1} and \eqref{eqB2}, yielding
\begin{eqnarray}
  \Gamma^{r}{}_{\theta \theta} &= \displaystyle\frac{\mathcal{A} }{g_{rr} \Phi' \frac{\dd\mathcal{A}}{\dd\Phi}}\left(\frac{r g'_{rr}}{2 g_{rr}}+g_{rr}-\frac{r g_{tt}'}{2 g_{tt}}-1\right)-\frac{r^2 \mathcal{V}}{\Phi' \frac{\dd\mathcal{A}}{\dd\Phi}}-\frac{r}{g_{rr}}\,,\label{gamma2}
\end{eqnarray}
which gives a connection component that depends on the coupling function, the potential, and also directly on the scalar field. In the following sections we will explore these two different approaches separately and find scalarised solutions for each of then. To do this, we will assume that the kinetic term coupling is always a constant, namely $\mathcal{B}(\Phi)=\beta$.

\subsection{Solutions assuming that the connection equation is satisfied independently of the scalar field and $\mathcal{V}=0$}\label{secsol1}
In this section, we will assume that the connection component does not depend on the scalar field, which leads to the solution~\eqref{gamma}, as discussed above. Substituting~~\eqref{gamma} into the other equations eliminates $\Gamma^{r}{}_{\theta \theta}$ and simplifies the remaining equations.
Let us now assume the conditions
\begin{eqnarray}
 g_{rr}=\frac{1}{g_{tt}}\,,\quad \mathcal{V}(\Phi)=0\,,\label{recipro}
\end{eqnarray}
and replace the above solution in the remaining equations. By doing this, one notices that if one subtracts Eq.~\eqref{eqB1} with \eqref{eqB2} one obtains
\begin{eqnarray}\label{condition}
g_{tt}(r)=\frac{4}{\left(2 \frac{\dd\mathcal{A}}{\dd\Phi}+r \beta \Phi'\right)^2}\Big(\frac{\dd\mathcal{A}}{\dd\Phi}\Big)^2\,,
\end{eqnarray}
where we have assumed that $2 \frac{\dd\mathcal{A}}{\dd\Phi}+r \beta \Phi'\neq 0$ since that case does not provide any solutions. Now, if we replace the above condition and its derivatives back to~\eqref{eqB1} and \eqref{eqB2} we find that only when $\Gamma^{r}{}_{\theta \theta}<0$, the system contains solutions (choosing the minus sign in~\eqref{gamma}). From~\eqref{eqB3} we find that there is a unique coupling function consistent with \eqref{gamma}, which is
\begin{equation}
\label{eq: BBMB A}
  \mathcal{A}(\Phi)=  C_1 (1-2 C_2 \Phi)^2\,,
\end{equation}
which from~\eqref{condition} means that the metric component behaves as
\begin{eqnarray}
 g_{tt}(r)=\frac{64 C_1^2 C_2^2 (-2 C_2 \Phi+1)^2}{\left(16 C_1 C_2^2 \Phi-8 C_1 C_2+\beta  r \Phi'\right)^2}\,.
\end{eqnarray}
Finally, to solve the system we replace all the above equations into~\eqref{eqB2} which gives us an implicit form of the scalar field satisfying (for $\beta +32 C_1 C_2^2\neq 0$)
\begin{eqnarray}\label{implicit}
 2 K_4 r \sqrt{\beta +32 C_1 C_2^2} (-2 C_2 \Phi+1)^{\frac{\beta }{32 C_1 C_2^2}+1} \left(\left(\beta +32 C_1 C_2^2\right) (-2 C_2 \Phi+1)^{\frac{\beta }{32 C_1 C_2^2}-1}+2 \beta  K_3\right)=1\,,
\end{eqnarray}
while for the case $\beta +32 C_1 C_2^2=0$, it is possible to analytically solve the equation explicitly for the scalar field giving us
\begin{eqnarray}
 \Phi(r)=\Phi_0\Big(1-\frac{M}{r}\Big)^{-1/2}+\frac{1}{2C_2}\,.\label{scalar1}
\end{eqnarray}
Thus, by assuming~\eqref{recipro} there are two solutions
for which the connection component $\Gamma^{r}{}_{\theta \theta}$ does not depend on the scalar field. The first one requires $\beta +32 C_1 C_2^2\neq 0$ and can be only implicitly written (see Eq.~\eqref{implicit}) since $\Phi$ cannot be analytically solved. The other solution has a scalar field given by~\eqref{scalar1} and the metric, coupling function and the connection component become
\begin{subequations}\label{solution1}
\begin{align}
\label{solution1metric}
ds^2&=-\Big(1-\frac{M}{r}\Big)^2\dd t^2+\Big(1-\frac{M}{r}\Big)^{-2}\dd r^2+r^2\dd \Omega^2\,,\quad \Gamma^{r}{}_{\theta \theta}=M-r\,,\\
 \mathcal{A}(\Phi)&=-\frac{\beta  \Phi ^2}{8}-\frac{\beta }{32 C_2^2}+\frac{\beta  \Phi }{8 C_2}\,,\quad \mathcal{V}(\Phi)=0\,.
 \end{align}
\end{subequations}

Despite a different initial look, the solution \eqref{scalar1}, \eqref{solution1} above is actually in accord with the observation made while studying the no-hair theorem, namely that quadratic coupling function with vanishing potential is amenable for yielding solutions. Indeed, a redefinition of the scalar field, $\Phi = \tilde{\Phi} + \tfrac{1}{2C_2}$, brings the coupling function into the quadratic form \eqref{eq: no-hair beta} while the kinetic function remains $\mathcal{B}(\tilde{\Phi})=\beta$ as in the no-hair discussion. After the redefinition the solution reads
\begin{align}
    \tilde{\Phi}(r)=\Phi_0\Big(1-\frac{M}{r}\Big)^{-1/2} \,, \qquad \mathcal{A}(\tilde{\Phi}) = -\frac{\beta \tilde{\Phi}^2}{8}\,,\label{scalar1simple}
\end{align}
with the metric and connection still given by \eqref{solution1metric}. The integration constant $\Phi_0$ gives the spatially asymptotic value of the scalar field, and can be interpreted as giving an inverse multiplicative modification to the effective gravitational constant in the spatial asymptotics. In contrast with some other solutions discussed later in the paper, the present configuration is characterized by a finite value of the asymptotic gravitational constant.

The metric \eqref{solution1metric} has the same form as an extremal Reissner–Nordstr\"{o}m black hole and it exhibits a horizon at $r=M$. In fact, the metric reproduces the well known BBMB black hole geometry in Riemannian conformal scalar-tensor theory~\cite{bocharova1970exact,Bekenstein:1974sf,Bekenstein:1975ts}, as well as the very recently found analogue in torsional scalar-tensor teleparallel gravity \cite{Bahamonde:2022lvh}. Several comparisons with these solutions can be made. In all three cases the scalar field is finite in the asymptotics and diverges at the horizon. Inside the horizon the BBMB scalar field becomes negative, while the teleparallel solutions show imaginary scalar field. However, the latter is not necessarily a fundamental flaw, since the nonminimal coupling function is quadratic and remains real. Interestingly, the form of the scalar field as well as of the nonminimal coupling function in the torsion and nonmetricity scalar-tensor models that possess these solutions is exactly the same, Eq.\ \eqref{scalar1simple}, which may hint for a closer correspondence between these particular models. Also, while the nonmetricity scalar of the solution \eqref{solution1metric}, \eqref{scalar1simple} is given by $Q=-2 M^2/r^4$, in the respective torsional scalar-tensor model the torsion scalar behaves rather similarly, $T=-2M^2/r^4+8/r^2$ \cite{Bahamonde:2022lvh}.

As a final remark here, let us note that in order to keep the sign of the effective gravitational constant positive outside of the horizon we need $\beta<0$. However, this gives the scalar field kinetic term an unnatural sign in the action, which in the minimal coupling case would signal a ghost instability. However, in the present case of nonminimally coupled scalar field the situation would require a more thorough analysis, which remains beyond the scope of this work (see in~\cite{Sezgin:1981xs,blagojevic1987extra,Lin:2018awc,BeltranJimenez:2019hrm,BeltranJimenez:2020sqf,Jimenez-Cano:2022sds} a list of relevant analyses on the stability in metric-affine geometry).

\subsection{Solutions assuming that the connection component $\Gamma^{r}{}_{\theta \theta}$ depends on the scalar field and $\mathcal{V}(\Phi)=0$}\label{secsol2}

This section will be devoted to studying the case where the connection component $\Gamma^{r}{}_{\theta \theta}$ is given by~\eqref{gamma2}, and the scalar potential is zero.

The first important remark about the zero potential case is that if we subtract~\eqref{eqB1} with~\eqref{eqB2} and use~\eqref{eqB3} we find that the following relationship must hold for any coupling function:
\begin{eqnarray}
 \Phi(r)&=&\mathcal{A}^{(-1)}\left(\frac{\tilde{\Phi}_0 \sqrt{g_{rr}g_{tt}}}{r^2 g_{tt}'}\right)\,,\label{eqphi}
 \end{eqnarray}
Then, if one assumes a model (a coupling $\mathcal{A}(\Phi)$) one can obtain the scalar field by taking the inverse of that function and then by evaluating this function in the argument appearing in the above equation. Note that the parenthesis above denotes the argument of the function.
 
 To find solutions, we will assume that the coupling function and the metric component satisfy
 \begin{eqnarray}
 \mathcal{A}(\Phi)=p\frac{\beta}{8} \Phi^2\,,\quad g_{rr}=\frac{1}{g_{tt}}\,,\label{conditions2}
 \end{eqnarray}
where $p$ is a constant. if we replace~\eqref{conditions2} into the~\eqref{eqB1} we find the following differential equation
\begin{align}
    r \Phi g_{tt}' \left(p \Phi+2 r \Phi'\right)+2 (p+1) r (g_{tt}-1) \Phi \Phi'+p (g_{tt}-1) \Phi^2+4 r^2 g_{tt} \Phi'^2=0\,.\label{transi}
\end{align}
Note that the case $\Phi (r)= \Phi_0 r^{-\frac{p}{2}}$ is a special case where $g'_{tt}$ vanishes from this equation and does not lead to any solution. Now, if we solve the above equation for $g'_{tt}$, one can rewrite~\eqref{eqB3} in the following form
\begin{eqnarray}
 \Big(\Phi \left(r \Phi''+\Phi'\right)-r \Phi'^2\Big) \Big(p^2 (g_{tt}-1) \Phi^2+4 p r g_{tt} \Phi \Phi'+4 r^2 g_{tt} \Phi'^2\Big)=0\,.\label{branching}
\end{eqnarray}
From this equation, there are two different branches depending on which bracket one solves. 

\subsubsection{Scalar field power law solutions}
By solving the first bracket of Eq.\ \eqref{branching}, we find that the scalar field must have a power-law form. By further replacing this in~\eqref{transi} one can easily solve for $g_{tt}$, and then one can solve the remaining field equations~\eqref{eqB4}-\eqref{eqB5} by constraining one of the integration constants appearing in the computation in term of $p$. All of this procedure provides the following solution for the system
\begin{align}
   ds^2&=-\Big(C_1^2-C_2 r^{1/C_1}\Big)\dd t^2+\Big(C_1^2-C_2 r^{1/C_1}\Big)^{-1}\dd r^2+r^2\dd \Omega^2\,,\quad  \Gamma^{r}{}_{\theta \theta}=-C_1 r\,,\\
   \Phi(r)&=\Phi_0 r^{-\frac{C_1+1}{2 C_1}}\,, \quad \mathcal{A}(\Phi)=\Big(\frac{C_1+1}{C_1-1}\Big)\frac{\beta  \Phi ^2}{8}\,,\quad \mathcal{V}(\Phi)=0\,.
   \end{align}
The case $C_1=+1$ is only possible if $\beta=0$ which leads to a particular solution considered later, Eq.~\eqref{solution0}. Moreover, if $C_1=-1$, the dynamics of the scalar field is not present since both the coupling function and the scalar field are zero. This is the only case where this solution is asymptotically Minkowski and further, it becomes a Schwarzschild solution with $C_2=2M$. For the other cases where $C_1\neq \pm 1$, the above solution is not asymptotically Minkowski but it can nevertheless represent a black hole solution with a horizon at $r=(C_1^2/C_2)^{C_1}$. 
Another remark is that the coupling function vanishes or diverges at infinity and then the effective gravitational constant would also diverge or vanish in that limit. For that reason the physical relevance of this class of solutions in not very clear. Let us note that although we may redefine the scalar field by adding a constant term to the solution like in Sec.\ \ref{secsol1}, that would also redefine the coupling function and the behavior of the effective gravitational constant would be the same.
   
\subsubsection{Solutions in terms of the Lambert function, $p=+1$}
Other solutions can be found by imposing that the second bracket in Eq.~\eqref{branching} is zero leading to the condition
\begin{eqnarray}
 \Phi'=-\frac{p \Phi  }{2 r}\left(1\pm \frac{1}{\sqrt{g_{tt}}}\right)\,.\label{phiprimegeneral}
\end{eqnarray}
If we now replace this expression in~\eqref{eqB1} we find the following differential equation
\begin{align}
    -r g_{tt}'\pm 2 p  \sqrt{g_{tt}}+(p-1) g_{tt}+p+1=0\,.\label{diffeq}
\end{align}
This equation can be easily solved for $p=-1$ which actually gives the same solution as the BBMB solution found previously (see Eq.~\eqref{solution1}) with $C_2\rightarrow \infty$. Furthermore, the case $p=+1$ can be straightforwardly solved yielding the following exact solution of the system~\eqref{eqB1}-\eqref{eqB5},
\begin{align}\label{solution2}
    ds^2&=-\left(1+W\left(\frac{-M}{ r}\right)\right)^2\dd t^2+\left(1+W\left(\frac{-M}{ r}\right)\right)^{-2}\dd r^2+r^2\dd \Omega^2\,,\quad \Gamma^{r}{}_{\theta \theta}=-r \left[1+W\left(-\frac{M}{r}\right)\right]\,,\\
  \Phi(r)&= \Phi_0 \left(-\frac{M}{r W\left(-\frac{M}{r}\right)} \right)^{1/2}\,,\quad \mathcal{A}(\Phi)=\frac{\beta  \Phi ^2}{8}\,,\quad \mathcal{V}(\Phi)=0\,, 
\end{align}
where $W(z)$ is the Lambert function defined as $W(z)e^{W(z)}=z$ and $M$ is a constant~\cite{corless1996lambertw}.
It is simple to check that $\Phi(r\rightarrow\infty)\rightarrow \Phi_0$, which provides $\mathcal{A}(\Phi(r\rightarrow\infty))=\frac{\beta \Phi_0^2}{8}$ and then $\Phi_0$ modifies the effective gravitational constant which remains finite. This solution is asymptotically Minkowski and represents a black hole solution with an event horizon located at $r=e M$ (if $M>0$). This means that the horizon is shifted from $2M$ to $e\,M$ with respect to the Schwarzschild solution. To understand how the parameters of solutions are related to the mass, it is useful to define the ADM mass which is
\begin{equation}\label{ADM}
    M_{\rm ADM}=\frac{1}{2}\lim_{r \to \infty}\left[r\left(g_{rr}-1\right)\right]\,.
\end{equation}
For the Lambert function~\eqref{solution2}, we find that $M_{\rm ADM}=M$, meaning that $M$ represents the mass of the black hole. One might also mention that this solution behaves similarly to the extremal Reissner–Nordstr\"{o}m black hole which is equivalent to the BBMB solution described by~\eqref{solution1}. The reason for this is that it can be also written in a squared form and one might argue that the above metric could come from a more general metric with $g_{tt}=1+2W\left(\frac{-M}{ r}\right)+W\left(\frac{-Q}{ r^2}\right)$ in the limit $M=Q$. That metric would be very similar to the Reissner–Nordstr\"{o}m where $M$ is the mass and $Q$ the charge but the Lambert function is operating on each term.

\begin{figure}[htp]
    \centering
    \includegraphics[scale=1.3]{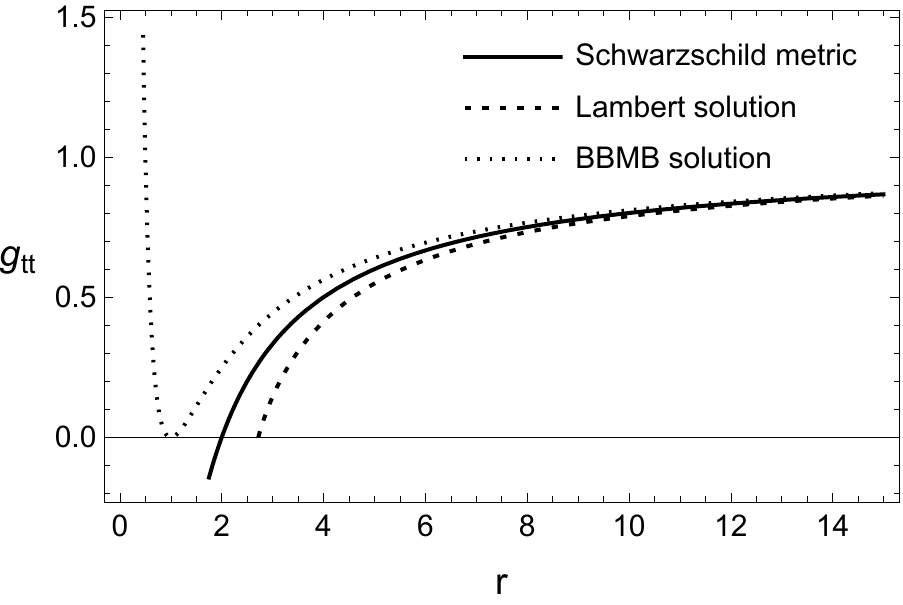}
    \caption{Plot of $g_{tt}$ versus $r$ for the Schwarzschild metric, the BBMB~\eqref{solution1} and the Lambert solution~\eqref{solution2} with $M=1$.}
    \label{fig1}
\end{figure}

If we now assume $M\ll 1$ and expand the above expressions up to second order in $M$, we find that the resulting metric exactly coincides with the Schwarzschild one with  nontrivial scalar field and connection components behaving as $\Phi(r)=\Phi_0\left( 1 - \tfrac{M}{2r}+\mathcal{O}(M^2) \right)$  and $\Gamma^{r}{}_{\theta \theta}=M-r+\mathcal{O}(M^2)$, respectively. The behaviour of the above solution can be seen in Fig.~\ref{fig1} where we plot and compare it with the Schwarzschild metric and also with the BBMB when $M=1$. We can also notice that there is an important difference between this solution and the Schwarzschild and BBMB ones, which is that $g_{tt}$ cannot describe the interior of the black hole since that the metric for $r<e M$ is not defined. 

Finally, let us note that the nonminimal coupling function $\mathcal{A}$ here is exactly in the same form as for the analogue of the BBMB solution, Eq.\ \eqref{scalar1simple}, except for the overall sign. As we expect the effective gravitational constant to be positive outside of the horizon, here the constant $\beta$ must be positive, while in the BBMB case it had to be negative. This leads to a speculative idea that the present solution with the Lambert function is somehow a counterpart to the BBMB solution when the sign of the kinetic term of the scalar field is flipped. Its analogues in the Riemannian or in the torsional teleparallel scalar-tensor theory have not been reported (up to our knowledge).

\subsubsection{More general classes of solutions}

For other values of $p\neq \pm 1$, the differential equation~\eqref{diffeq} leads to the following implicit form of the metric component
\begin{eqnarray}
    \left( 1 \pm \frac{1}{\sqrt{g_{rr}}}\right)\left(1\mp \frac{1}{K\sqrt{g_{{rr}}}}\right)^{K}=\frac{C_2}{r}\,,\quad K=\frac{1+p}{1-p}\,.\label{master2}
\end{eqnarray}
This algebraic equation can be explicitly solved for some specific values of $p$, yielding different exact solutions for the system. One remark about the above equation is that since they satisfy~\eqref{diffeq}, the nonmetricity scalar for any $p$ can be written in the following compact form
\begin{eqnarray}\label{nonmQ}
 Q=\frac{2 \left(\sqrt{g_{tt}}\pm 1  \right)^2 \left(p \sqrt{g_{tt}}\pm (p+1)  \right)}{r^2 \sqrt{g_{tt}}}\,,
\end{eqnarray}
where we can easily see that for the minus sign, this scalar will always become zero in the Minkowski limit ($g_{tt}\rightarrow 1$), but for the plus sign, the only way to have $Q=0$ in that limit is when $p=-1/2$. Thus, for $p\neq -1/2$, it is better to choose the solutions with the minus sign above such that we have that $Q=0$ for Minkowski. 

For the particular cases of $p=\{\pm2,\pm 3\}$ and by redefining $C_2$ such that one can always associate this constant to the ADM mass by~\eqref{ADM}, we can find different exact solutions of the system~\eqref{eqB1}-\eqref{eqB5}. In the next subsections, we will show their explicit forms. Note that in all of these solutions when $r\rightarrow \infty$, the scalar field vanishes and then $\mathcal{A}$ is also zero, meaning that the effective gravitational constant for these solutions diverges at infinity. It is also important to notice that the differential equation which connects the metric with the scalar field~\eqref{phiprimegeneral} holds for any value of $p$.

\subsubsection{Scalar field square root solution with $p=+2$}
The first simplest case where one can solve~\eqref{master2} is for $p=+2$ which gives the following exact solution
\begin{align}\label{solution3}
    ds^2&=-9\Big(1-\frac{8 r}{3 M \sqrt[3]{H_1(r)}}+\frac{1}{9} \sqrt[3]{H_1(r)}\Big)^{2}\dd t^2+\frac{1}{9}\Big(1-\frac{8 r}{3 M \sqrt[3]{H_1(r)}}+\frac{1}{9} \sqrt[3]{H_1(r)}\Big)^{-2}\dd r^2+r^2\dd \Omega^2
    \,,\\
    \Gamma^{r}{}_{\theta \theta}&=\frac{r \left(-r g_{tt}(r) g_{tt}''(r)+r g_{tt}'(r)^2-(g_{tt}(r)+1) g_{tt}'(r)\right)}{r g_{tt}''(r)+2 g_{tt}'(r)}\,,\\
  \Phi(r)&=\Big[\frac{108 M^2\Phi_0 H_1(r)^{5/3}}{\beta  r^2 \left(-9 M \sqrt[3]{H_1(r)}-M H_1(r)^{2/3}+24 r\right) \left(M H_1(r)^{2/3} H_1'(r)+24 r H_1'(r)-72 H_1(r)\right)}\Big]^{1/2}\,,\\
  \mathcal{A}(\Phi)&=\frac{\beta  \Phi ^2}{4}\,,\quad \mathcal{V}(\Phi)=0\,,
\end{align}
where $H_1(r)=24 r \left(\sqrt{81 M+24 r}-9 \sqrt{M}\right)/M^{3/2}$. We did not show the explicit form of $\Gamma^{r}{}_{\theta \theta}$ since it becomes cumbersome by substituting $g_{tt}$. This metric is asymptotically Minkowski if $M>0$ and contains one horizon. This means that this solution cannot describe an asymptotically Minkowski black hole solution since the ADM mass~\eqref{ADM} is $-M$. The scalar field also vanishes asymptotically, and the effective gravitational constant becomes infinite there.

 \subsubsection{Scalar field square root solution with $p=-2$}
Another solution can be found by setting $p=-2$ in ~\eqref{master2} yielding
\begin{align}
    ds^2&=-\Big(1+\frac{M^2}{2r \sqrt[3]{H_2(r)} }-\frac{\sqrt[3]{H_2(r)} M}{r^2}\Big)^{2}\dd t^2+\Big(1+\frac{M^2}{2r \sqrt[3]{H_2(r)} }-\frac{\sqrt[3]{H_2(r)} M}{r^2}\Big)^{-2}\dd r^2+r^2\dd \Omega^2\,,
    \\
    \Gamma^{r}{}_{\theta \theta}&=\Big[\frac{64 H_2(r)^{5/3} r^3 \Phi_0 \left(4 H_2(r)-2 r^3\right)\left(M^2 r-2 H_2(r)^{2/3} M+2 \sqrt[3]{H_2(r)} r^2\right)^{-1}}{\beta  M  \left(2 H_2(r)^{2/3} r^3 \left(24 H_2(r)+M^3\right)-64 H_2(r)^{8/3}+16 H_2(r)^2 M r+M^4 r^4\right)}\Big]^{1/2}
    \,,\\
  \Phi(r)&=\Big[\frac{8 M r^4 \Phi_0}{\beta  \left(H_2(r)^{2/3} \left(3 M^3 r-8 r^4\right)+6 H_2(r)^{4/3} M^2+8 H_2(r)^{5/3} r-8 \sqrt[3]{H_2(r)} M^2 r^3-3 M^4 r^2\right)}\Big]^{1/2}\,,\\
  \mathcal{A}(\Phi)&=-\frac{\beta  \Phi ^2}{4}\,,\quad \mathcal{V}(\Phi)=0\,,
\end{align}
with $H_2(r)=\frac{1}{4} r^2 \left(M \sqrt{\frac{4 r^2}{M^2}+\frac{2 M}{r}}+2 r\right)$. This solution has a ADM mass given by $M$, and it is  asymptotically Minkowski when this mass is positive. Further, when $M\rightarrow 0$ we recover a Minkowski metric. However, for $M>0$, the above metric cannot describe a black hole solution since there are not event horizons. The only way to obtain an event horizon is to choose $M<0$ which indeed provides a black hole solution but is not asymptotically Minkowski and its ADM mass would be a negative quantity. The scalar field vanishes asymptotically, and the effective gravitational constant becomes infinite there.

 \subsubsection{Scalar field square root solution with $p=+3$}
The case $p=3$ can be also solved in~\eqref{master2} giving us
\begin{align}\label{solution6a}
   ds^2&=-4 \left(1-\frac{r }{4 M}\left(1-\frac{\epsilon}{\eta }H_3(r)\right)\right)^2\dd t^2+\frac{1}{4}\left(1-\frac{r }{4 M}\left(1-\frac{\epsilon}{\eta }H_3(r)\right)\right)^{-2}\dd r^2+r^2\dd \Omega^2\,,\\
   \Gamma^{r}{}_{\theta \theta}&=\frac{r \left(\epsilon  \left(-4 M^3+13 M^2 r-7 M r^2+r^3\right)-\eta  r H_3(r) \left(5 M^2-5 M r+r^2\right)\right)}{M \left(\eta  r H_3(r) (2 M-r)+\epsilon  \left(2 M^2-4 M r+r^2\right)\right)}\,,\\
    \Phi(r)& =\Phi_0 \Big[\frac{-4 M^3 H_3(r)}{r (r \epsilon  H_3(r)+4 \eta  M-\eta  r) (\eta  r H_3(r)+2 M \epsilon -r \epsilon )}\Big]^{1/2} \,,\\
    \mathcal{A}(\Phi)&=3\frac{\beta  \Phi ^2}{8}\,,\quad \mathcal{V}(\Phi)=0\,,
   \end{align}
where $H_3(r)=\sqrt{1-\frac{4 M}{r}}$, $\epsilon=\pm 1$ and $\eta=\pm 1$. When $\epsilon/\eta=1$ this solution represents a scalarised asymptotically Minkowski black hole configuration whose mass is given by $M$ and it contains one horizon  located at $r=4M$. Further, the Minkowski limit is obtained smoothly when $M=0$ and $Q\rightarrow 0$ in that limit. At $r\rightarrow \infty$ the scalar field becomes $\Phi_0$ which is a condition that is expected for a well-behaved solution (and then this constant acts as a finite modifier of the effective gravitational constant). Regarding the singularities, the Levi-Civita Kretschmann invariant diverges not only at $r=0$ but also at the horizon $r=4M$. When $\epsilon/\eta=-1$, this metric is not asymptotically Minkowski but still the metric contains one event horizon at $r=4M$.

\subsubsection{Scalar field square root solution with $p=-3$}
   The last solution that we will explicitely show is when $p=-3$, which from~\eqref{master2} gives us
\begin{align}
   ds^2&=\frac{1}{r^4}\left(M^2+r^2\right) \left(M+\epsilon\sqrt{M^2+r^2}\right)^2\dd t^2-r^4\left(M^2+r^2\right)^{-1} \left(M+\epsilon\sqrt{M^2+r^2}\right)^{-2}\dd r^2+r^2\dd \Omega^2\,,\\
   \Gamma^{r}{}_{\theta \theta}&=-r-\frac{M^2}{r}-\frac{\epsilon  }{r}\left(M \sqrt{M^2+r^2}\right)\,,
   \\
    \Phi(r)&=\Phi_0 \Big[\frac{r^3\epsilon}{ \left(\epsilon  \sqrt{M^2+r^2}+M\right) \left(2 M \left(\epsilon  \sqrt{M^2+r^2}+M\right)+r^2\right)}\Big]^{1/2}\,,\\
    \mathcal{A}(\Phi)&=-3\frac{\beta  \Phi ^2}{8}\,,\quad \mathcal{V}(\Phi)=0\,,\label{solution6}
   \end{align}
  This solution is always asymptotically Minkowski and $\Phi(r\rightarrow \infty)\rightarrow \Phi_0$, but it does not contain horizons, so that it cannot describe a black hole solution. Its ADM mass is described by $\epsilon M$. 
  
\subsubsection{Remark about other solutions}
There are other solutions for different $p$ where one can solve~\eqref{master2}, such that when $p=\{\pm 1/3,\pm 1/2\}$ but they are similar to the above solutions so we will omit them here. Notice that all the solutions can be written as square binomial terms in the metric.  The nonmetricity scalar~\eqref{nonmQ} is vanishing for all of the solutions found in their respective Minkowski limit. 

\subsection{Solutions assuming $\mathcal{B}(\Phi)=\beta=0$}

\subsubsection{Without a potential}
In this section we will study the case where the kinetic term vanishes $\mathcal{B}(\Phi)=\beta=0$. If we assume this and further $\mathcal{V}(\Phi)=0$ then the scalar field equation implies either the minimally coupled case $\mathcal{A}'=0$ or $Q=0$. In the latter case there is a generic metric solution of the system
\begin{eqnarray}\label{solution0}
 ds^2&=-\Big(1-2M r^{1/C_1}\Big)\dd t^2+\Big(1-2M r^{1/C_1}\Big)^{-1}\dd r^2+r^2\dd \Omega^2\,,
\end{eqnarray}
with $\Gamma^{r}{}_{\theta \theta}=-r$, as long as the following relationship holds
\begin{align}\label{caseeee}
    \Phi '=-\frac{(C_1^{-1}+1) \mathcal{A}(\Phi)}{r \frac{\dd\mathcal{A}}{\dd\Phi}}\,.
\end{align}
This generic solution has $Q=0$ but still the dynamics of the metric depends on the scalar field (even though we eliminated the kinetic and potential term). As long as $C_1<0$, this solution is asymptotically Minkowski and has one horizon located at $r=(2M)^{-C_1}$ as long as $M$ is positive. If one computes the ADM mass~\eqref{ADM}, we find that this mass diverges except for $-1 \leq C_1< 0$. Furthermore, the case $-1 < C_1< 0$ gives a zero ADM mass while the particular case where $C_1\rightarrow -1$ gives $M$ but only provides us a trivial scalar field with the metric being the Schwarzschild one. For completeness, let us consider an explicit example satisfying~\eqref{caseeee}. If we take $\mathcal{A}(\Phi)= \mathcal{A}_0\Phi^l$, we find that the scalar field must behave as $\Phi (r)= \Phi_1r^{-\frac{C_1+1}{C_1 l}}$. For this particular case the scalar field is asymptotically Minkowski for $(l<0\land -1<C_1<0)\lor (l>0\land C_1<-1)$, but $\mathcal{A}(\Phi(r\rightarrow\infty))\rightarrow 0$.

\subsubsection{$f(Q)$ gravity}

As discussed in Sec.~\ref{sec:symmetricTG}, the scalar-tensor theory that we are studying contains $f(Q)$ gravity represented by a scalar field with vanishing kinetic term, $\mathcal{B}=0$, while the other model functions $\mathcal{A}$ and $\mathcal{V}$ related to the function $f(Q)$. It is straightforward to substitute $f(Q)$ into the equations \eqref{eqB1}--\eqref{eqB5} and study the $f(Q)$ theory in the respective scalar tensor representation. This can be done by considering the choices expressed in~\eqref{transftofQ} where the scalar field $\Phi$ is equivalent to the nonmetricity scalar $Q$. 

We can readily check that the example solution found in Ref.\ \cite{DAmbrosio:2021zpm}, for a power-law $f(Q) = Q^\kappa$ theory in the $c=k=0$ branch of connections, does satisfy the scalar-tensor field equations as well. Transforming this theory into scalar-nonmetricity, one gets
\begin{align}
    \mathcal{A}(\Phi) = \kappa \Phi^{\kappa -1} \,, && \mathcal{B}(\Phi) = 0\,, && \mathcal{V}(\Phi) = \frac{1}{2}\Phi^\kappa(\kappa-1)
\end{align}
with $Q = \Phi$. Using these transformations, one notices that the modified Klein Gordon equation~\eqref{eqB5} is instantaneously satisfied. To obtain a solution, one can assume that 
\begin{equation}
    \Gamma^r_{\ \theta \theta} = -\lambda r\,.
\end{equation}
The case $\lambda=+1$ is a limiting case where there are only trivial solutions. An interesting solution is obtained by setting
\begin{align}
 \lambda = \frac{5-8\kappa +4\kappa^2}{5-14\kappa + 8\kappa^2}
\end{align}
(where $\kappa \neq \tfrac{1}{2}, \tfrac{5}{4}$)
yielding
\begin{align}
    ds^2&= -c_1 r^\beta \Big(1 - \frac{2M}{r^{\beta-\alpha}} \Big)\dd t^2+  \frac{C}{1-\frac{2M}{r^{\beta-\alpha}}} \dd r^2+r^2\dd \Omega^2\,,
\end{align}
where the constants must satisfy
\begin{align}
    \beta = \frac{8(2\kappa-3)(\kappa-1)\kappa}{5+4(\kappa-2)\kappa}&& \alpha = \frac{(2\kappa-3)(5+4\kappa(2\kappa-3))}{5+4(\kappa-2)\kappa} && \gamma = \alpha - \beta && C = \left(\frac{8\kappa^2-14\kappa +5}{4\kappa^2-8\kappa +5} \right)^2\,.
\end{align}
The nonmetricity scalar for this solution behaves as \begin{align}
    Q = \Phi & = -\frac{8 \kappa^2(2 \kappa-3)^2}{r^2 (2 \kappa -1)^2(4\kappa-5)}\,.
\end{align}
Clearly, when $\kappa = 1$ the metric becomes the Schwarzschild one and $f(Q)=Q$, which is GR, while the scalar field drops out of the action and equations completely. 

Let us finish this section by obtaining an equation that must be valid for any arbitrary function $f(Q)$. If we solve~\eqref{eqB1} and \eqref{eqB2} for $f_Q$ and $f_{QQ}$ and then, we replace those expressions in~\eqref{eqB3} we find that 
\begin{eqnarray}
 0&=&r^2 g_{tt} g_{rr}' \left(r g_{tt}'-2 g_{tt}\right)+g_{rr} \left[2 g_{tt}^2 \left((\Gamma^r{}_{\theta \theta})^2 g_{rr}'+2 \Gamma^r{}_{\theta \theta}+2 r\right)+r^3 g_{tt}'^2-2 r^2 g_{tt} \left((\Gamma^r{}_{\theta \theta}+r) g_{tt}''+g_{tt}'\right)\right]\nonumber\\
 &&+2 g_{rr}^2 g_{tt} \left((\Gamma^r{}_{\theta \theta})^2 g_{tt}'-2 g_{tt} (\Gamma^r{}_{\theta \theta}+r)\right)\,.
\end{eqnarray}
This equation must be true for any function $f(Q)$, meaning that it is an universal relationship that must always hold in this theory. This equation is similar to the ones derived in~\cite{Golovnev:2021htv,Bahamonde:2021srr} for $f(T)$ gravity case and as pointed out in those references, it was a pivotal point to find exact solutions in the theory. From the above equation is simpler to see that if $g_{rr}=1/g_{tt}$, the only solutions are either $\Gamma^r{}_{\theta \theta}=-r$ or the (Anti)de-Sitter–Schwarzschild. The first condition in the end also restricts the form of the metric to be (Anti)de-Sitter–Schwarzschild and $f=Q+\textrm{const}.$ In the following section \ref{sec:theorem} we will prove this feature in more detail.

\subsection{Solution assuming a nonvanishing potential}
As mentioned above, finding analytic  solutions in the general case is difficult since in the equations all quantities are mixed together and in order to disentangle them one needs to make some assumptions. When there is a nonvanishing potential and also $\beta\neq0$, the field equations are much harder to treat since the Eq.~\eqref{eqphi} is no longer valid. We may still rely on the fact that the no-hair theorem (see Secs.~\ref{sec:hair} and \ref{secondbranch})  establishes that a quartic potential with quadratic coupling might lead to exact analytic solutions. The resulting equations are not that simple, but it can be found that by assuming that the connection component and the scalar field behave as
\begin{align}
    \Gamma^r{}_{\theta\theta} = - r \,, \qquad \Phi = \frac{\Phi_0}{r}\,,
\end{align}
it is possible to find the following exact solution to the field equations:
\begin{align}\label{solution2B}
    ds^2&=-c_1 r^4 \left(1-\frac{2M}{r}\right)\dd t^2+\Big(1-\frac{2M}{r}\Big)^{-1}\dd r^2+r^2\dd \Omega^2\,,\quad  \mathcal{A}(\Phi)=\frac{\beta \Phi^2}{4} \,, \quad \mathcal{B}(\Phi) = \beta \,, \quad \mathcal{V}(\Phi) = - \frac{\beta \Phi^4}{2 \Phi_0^2} \,,
\end{align}
where $c_1, \Phi_0$ are constants. The metric has a horizon at $r=2M$. This solution is not asymptotically Minkowski, but the boundary term of \eqref{trace of field equations integrated} still vanishes as $\mathcal{A} \to 0$ in the spatial asymptotics. Perhaps, we can realize now why many analytic solutions we found previously also have vanishing $\mathcal{A}$ in the asymptotics. As a constant kinetic function, quadratic coupling, and quartic potential solve Eq.\ \eqref{eq: no-hair beta} identically, the boundary integral on the RHS of \eqref{trace of field equations integrated} must vanish. The latter typically happens either if the asymptotics is Minkowski, or if $\mathcal{A}$ drops to zero asymptotically. Since $\mathcal{A}$ modifies the effective gravitational constant, the physical relevance of these second type of solutions is not immediately obvious.
 
\section{The case of $f(Q)$ gravity under the reciprocal condition $g_{rr}=1/g_{tt}$}\label{sec:theorem} 
In this section, we are going to analyse the particular case of $f(Q)$ gravity when the metric functions satisfy $g_{rr}=1/g_{tt}$. An interesting property happens by just assuming that condition. As discussed in previous sections, there are two sets of connections in spherical symmetry and only set 2 can have different solutions to GR. Within this set, there are two subcases (branches). In previous sections, we have discarded the branch one (see Sec.~\ref{sec:firstbranch} to understand the reason why we did this), but in this section we will also analyse that branch. As mentioned in the previous section, the scalar-tensor theory studied is reduced to $f(Q)$ gravity when Eq.~\eqref{transftofQ} holds. For simplicity in the notation, let us now use the field equations of $f(Q)$ gravity given by~\eqref{MetricFieldEqf(Q)}-\eqref{ConnectionFieldEqf(Q)}, instead of the scalar-tensor representation of $f(Q)$ gravity. 

We will then show the following theorem in $f(Q)$ gravity:
\begin{theorem}
    For $f(Q)$ gravity where the metric and connection are static and spherically symmetric, the metric functions $g_{tt}(r)$ and $g_{rr}(r)$ take on the reciprocal of each other if and only if the model coincides with STEGR + Constant. In addition, the solution in this case is provided by the (anti)de-Sitter–Schwarzschild geometry.
\end{theorem}

The demonstration of this theorem is as follows. For the first branch (within set 2), the vacuum $f(Q)$ gravity metric field equations described by~\eqref{MetricFieldEqf(Q)} when $g_{rr}=1/g_{tt}$ are given by
\begin{eqnarray}
 0&=&f_{QQ} Q' \left(-4 c^2 r^2 (k-2 c)^2 (\Gamma^r{}_{\theta\theta})^2+r g_{tt}^2 \left(16 c (2 c-k) \Gamma^r{}_{\theta\theta}+r (k-4 c)^2\right)+8 c (2 c-k) g_{tt} (\Gamma^r{}_{\theta\theta})^2\right)\nonumber\\
 &&+4 c r^2 (k-2 c) g_{tt} \Gamma^r{}_{\theta\theta} f+4 c (2 c-k) f_{Q} g_{tt} \Gamma^r{}_{\theta\theta} \left(2 r g_{tt}'+2 g_{tt}+r^2 Q-2\right)\,,\label{fQ1}\\
  0&=&f_{QQ} Q' \left(-4 c^2 r^2 (k-2 c)^2 (\Gamma^r{}_{\theta\theta})^2+r^2 (k-4 c)^2 g_{tt}^2+8 c (k-2 c) g_{tt} (\Gamma^r{}_{\theta\theta})^2\right)+4 c r^2 (2 c-k) g_{tt} \Gamma^r{}_{\theta\theta} f\nonumber\\
  &&-4 c (2 c-k) f_{Q} g_{tt} \Gamma^r{}_{\theta\theta} \left(2 r g_{tt}'+2 g_{tt}+r^2 Q-2\right)\,,\label{fQ2}\\
   0&=&f_{QQ} Q' \left(4 c^2 r (k-2 c)^2 (\Gamma^r{}_{\theta\theta})^2+4 c r (2 c-k) g_{tt} \Gamma^r{}_{\theta\theta} g_{tt}'+g_{tt}^2 \left(8 c (2 c-k) \Gamma^r{}_{\theta\theta}+r (k-4 c)^2\right)\right)\nonumber\\
   &&+4 c r (k-2 c) g_{tt} \Gamma^r{}_{\theta\theta} f+4 c (2 c-k) f_{Q} g_{tt} \Gamma^r{}_{\theta\theta} \left(r g_{tt}''+2 g_{tt}'+r Q\right)\,,\label{fQ3}
\end{eqnarray}
whereas the connection equation~\eqref{ConnectionFieldEqf(Q)} becomes
\begin{align}
   0&=\frac{4 c^2 r^2 (k-2 c)^2 (\Gamma^r{}_{\theta\theta})^2 f'_{Q} g_{tt}'}{g_{tt}^2}+8 c (k-2 c) (\Gamma^r{}_{\theta\theta})^2 f''_{Q}+f'_{Q} \left(r^2 (k-4 c)^2 g_{tt}'+16 c (k-2 c) \Gamma^r{}_{\theta\theta} (\Gamma^r{}_{\theta\theta})'\right)\nonumber\\
   &+r (k-4 c)^2 g_{tt} \left(r f''_{Q}+2 f'_{Q}\right)-  \frac{4 c^2 r (k-2 c)^2 \Gamma^r{}_{\theta\theta} \left(2 r f'_{Q} (\Gamma^r{}_{\theta\theta})'+\Gamma^r{}_{\theta\theta} \left(r f''_{Q}+2 f'_{Q}\right)\right)}{g_{tt}}\label{fQ4}\,.
\end{align}
We notice that if we solve~\eqref{fQ1} and~\eqref{fQ2} for $f_{QQ}$ and $f_Q$ and then we replace those expressions into~\eqref{fQ3} we find the following equation:
\begin{equation}
   0= c (2 c-k) g_{tt}^2 (\Gamma^r{}_{\theta\theta})^2 f \left(r^2 g_{tt}''-2 g_{tt}+2\right)\,.
\end{equation}
Since in this branch, $c\neq0, 2 c-k\neq0$, the only nontrivial solution to this equation is by imposing that the above parenthesis is zero which gives us the following form of the metric
\begin{equation}
    g_{tt}(r)=1-\frac{2M}{r}-\Lambda r^2\,,
\end{equation}
which is the (Anti)de-Sitter–Schwarzschild solution. This means that only by imposing $g_{rr}=1/g_{tt}$, the metric is always fixed to have this form. If we further replace this condition into~\eqref{fQ2} and~\eqref{fQ3} we respectively find
\begin{eqnarray}
 0&=&\left[-4 c^2 r^3 (k-2 c)^2 (\Gamma^r{}_{\theta\theta})^2+8 c (2 c-k) \Gamma^r{}_{\theta\theta} \left(2 M+\Lambda  r^3-r\right)^2+r (k-4 c)^2 \left(2 M+\Lambda  r^3-r\right)^2\right] (f-f_Q (Q-6\Lambda))\,,\nonumber\\
&&\\ 
  0&=&c (2 c-k) \Gamma^r{}_{\theta\theta} \left[\Gamma^r{}_{\theta\theta} \left(r^3 \left(-2 c^2+c k-\Lambda \right)-2 M+r\right)+(3 M-r) \left(2 M-r \left( r^2+1-\Lambda \right)\right)\right] (f-f_Q (Q-6\Lambda))\,,
\end{eqnarray}
where we have also used Eq.~\eqref{fQ1}. From the above equation, we notice that the only way to solve the system is by imposing $f-f_Q (Q-6\Lambda)=0$ which gives $f(Q)=Q/2-3\Lambda$. This calculation showed that in the first branch, the constrain $g_{rr}=1/g_{tt}$ restricts the form of the metric to be (Anti)de-Sitter–Schwarzschild and the form of the Lagrangian as STEGR plus a cosmological constant. 

Let us now study the same for branch 2 (within set 2) assuming the same condition $g_{rr}=1/g_{tt}$.
By subtracting \eqref{eqB1} from \eqref{eqB2} and using~\eqref{transftofQ} we find
\begin{align}
    \frac{2 (r+\Gamma^r{}_{\theta\theta})g_{tt} Q'}{r \Gamma^r{}_{\theta\theta}} f_{QQ} = 0 \,.
\end{align}
If $\Gamma^r{}_{\theta\theta}\neq -r$, the above equation is satisfied by either $f_{QQ} = 0$ or $Q=\textrm{const}.$ which in any of the two cases imply that the theory cannot go beyond STEGR (or GR) with a cosmological constant. Then, the only nontrivial option at this point is $\Gamma^r{}_{\theta\theta} = -r$, but that reduces the remaining part of the system \eqref{eqB1}-\eqref{eqB5} (with \eqref{transftofQ}) to
\begin{align}
     Q f_{QQ}= 0 \,,
\end{align}
which would again imply a vanishing scalar field or $f$ being linear in $Q$. Thus we can conclude that there are no nontrivial spherically symmetric static  solutions in the $c=k=0$ branch of $f(Q)$ gravity which are characterised by $g_{rr}=1/g_{tt}$. 

We just showed that for set 2 (for the two branches), the condition $g_{rr}=1/g_{tt}$ immediately constrains the theory and one cannot obtain solutions beyond GR (plus a cosmological constant). Since set 1 one also has the same result (trivially for any metric, see the discussion around Eq.\ \eqref{eq: tr set 1}), and we know that the unique solution of STEGR+Constant is the (anti)de-Sitter–Schwarzschild geometry, the theorem is already shown. It is interesting to mention that this theorem is analogous to the one proven in~\cite{Bahamonde:2021srr} for $f(T)$ gravity. This also shows another similarity between those two theories.

\section{Conclusions}\label{sec:conclusions}

In this work, we studied static spherical symmetric configurations in a symmetric teleparallel (zero torsion and curvature) scalar-tensor theory which contains a nonminimal coupling between the nonmetricity scalar and the scalar field. This theory reduces to the standard Einstein minimally coupled scalar-tensor theory when this coupling function is a constant. In the symmetric teleparallel framework, there are two different sets having the property that both the metric and the connection are spherically symmetric \cite{DAmbrosio:2021zpm}. While the first set does not provide anything new beyond the standard minimally coupled case, the second set provides field equations that are nontrivial and different to the standard Riemannian case. This set is further divided into two branches that have different field equations and then different dynamics. 

The main aim of this paper was to find exact asymptotically Minkowski spherically symmetric solutions and study if they can describe hairy black holes. First, in Sec.~\ref{sec:hair} and~\ref{secondbranch} we formulated a no-hair theorem for our theory in order to find possible routes to tackle the equations. We noticed that it is difficult to evade the no-hair theorem for the first branch of set two which also has more complicated field equations. Thus, we mainly focused on the second branch of set two where we were able to find several exact solutions depending on the choice of the coupling function and the potential.

After carefully studying the field equations, we split the analysis in different cases. The simplest nontrivial case that can be solved is obtained by assuming that there is no potential and the coupling function has a quadratic form $\mathcal{A}(\Phi)=\mathcal{A}_0 \Phi^2$. Depending on the proportionality constant, we found different exact solutions, all of them have the metric as $ds^2=-\tilde{g}_{tt}^2dt^2+1/\tilde{g}_{tt}^{2}dr^2+r^2d\Omega^2$, meaning that it is always possible to write it with square binomial terms. The first interesting solution is the BBMB (see Eq.~\eqref{solution1}) which coincides with an extremal Reissner–Nordstr\"{o}m metric. It is worth mentioning that this solution not only appears in the Riemannian case (in a conformal scalar-tensor theory), but also in a metric teleparallel scalar-tensor theory with torsion~\cite{Bahamonde:2022lvh}. We also found an asymptotically Minkowski  exact spherically symmetric solution containing a Lambert function (see~\eqref{solution2}) that, to the best of our knowledge, has not been reported in any scalar-tensor theory. This solution cannot describe the interior of a black hole but it might happen that it belongs to a more general solution containing two Lambert functions and then, one might describe the interior of a black hole. We also found other solutions with different values of $\mathcal{A}_0$ which gives rise to complicated forms of the metric containing roots terms (see Eqs~\eqref{solution3}-\eqref{solution6}). Using the definition of the ADM mass, we were able to distinguish between solutions that can describe hairy black holes or not. For example, the solution~\eqref{solution6a} can describe an asymptotically Minkowski hairy black hole solution with the scalar field having a nontrivial profile.

Other types of solutions were obtained when the kinetic term in the action vanishes. For example, Eq.~\eqref{solution0} provides a solution for the vanishing potential case and also the $f(Q)$ case was analysed (which is part of this limit). As already reported in~\cite{DAmbrosio:2021zpm}, there is an exact black hole solution for a power-law $f(Q)$ case. We further showed a theorem in Sec.~\ref{sec:theorem} which states that there are only trivial solutions in $f(Q)$ gravity if the condition $g_{rr}=1/g_{tt}$ holds. Interestingly, an analogous result was also shown in~\cite{Bahamonde:2021srr} for $f(T)$ gravity. Furthermore, it is also worth mentioning that some solutions presented in the torsional teleparallel case also appear in the symmetric teleparallel case (see~\cite{Bahamonde:2022lvh}). It would be interesting to clearly understand how these two different theories are connected (in spherical symmetry) and why one can find similar solutions.

One important aspect to mention about our work is that even though the Riemannian scalar-tensor theory constructed by couplings between the Levi-Civita Ricci scalar and the scalar field has been studied by many authors, it has been difficult to obtain exact asymptotically Minkowski hairy black hole solutions. Furthermore, there is a no-hair theorem in this theory which says that when there is no potential, there are not asymptotically Minkowski, static, non-extremal and spherically symmetric solutions beyond the Schwarzschild one~\cite{Zannias:1994jf,Saa:1996aw,Saa:1996qq}. It should be noted that the BBMB is an exact solution of this theory but this is an extremal solution that has certain pathologies such that being unstable under linear perturbations or the energy-momentum tensor of the scalar field diverges. However, since our theory is different to the conformal Riemannian theory, one would need to study its perturbations to fully understand if those instabilities appear in our theory or not. A priori, if we have the same solution at the background level, one cannot conclude that we will have the same behaviour at the perturbation one, meaning that for example, the quasinormal modes of our solution could be different to the standard Riemannian case (since the theories are different).

In addition, the extension to axially symmetric space-times could be an interesting route to analyse. To do this, first one would need to obtain the corresponding symmetric teleparallel connection preserving axial symmetry with the fulfilment of Eq.~\eqref{LieD_mag}. This would lead to more realistic solutions endowed with rotations, such as the Kerr solution or further modifications. A preliminary study for the case of torsional teleparallel gravity was recently adressed in \cite{Bahamonde:2020snl}, whereas for more general metric-affine geometries different types of solutions have been found~\cite{Bakler:1988nq,Obukhov:2019fti,Bahamonde:2021qjk}. 

Finally, it is worth mentioning that by omitting the solutions with a potential, the scalarised solutions found in our paper are of secondary type. This means that the scalar hair is not independent but is expressed in term of the black hole mass. In Riemannian geometry, there are certain scalar-tensor theories such as Einstein-Gauss-Bonnet~\cite{Antoniou:2017acq} where the scalar charge and the black hole mass are independent. Furthermore, in those theories, one can have spontenous scalarization triggered by the new corrections coming from the Gauss-Bonnet invariant~\cite{Silva:2017uqg,Antoniou:2017acq,Doneva:2017bvd,Herdeiro:2018wub,Silva:2018qhn}. These studies will be reported in the future in detail for symmetric and torsional teleparallel gravity.

\bigskip
\bigskip
\noindent
\section*{Acknowledgements}
S.B. is supported by JSPS Postdoctoral Fellowships for Research in Japan and KAKENHI Grant-in-Aid for Scientific Research No. JP21F21789. S.B. also acknowledges the Estonian Research Council grant PRG356 ``Gauge Gravity". J.G.V. was supported by the European Regional Development Fund and the programme Mobilitas Pluss (Grant No. MOBJD541). L.J.\ is supported by the Estonian Research Council grant PRG356 ``Gauge Gravity". The authors also acknowledge support by the European Regional Development Fund through the Center of Excellence TK133 ``The Dark Side of the Universe".

\bibliographystyle{utphys}
\bibliography{references}

\end{document}